\documentclass[aps,prd,onecolumn,superscriptaddress,nofootinbib]{revtex4-2}
\usepackage[ansinew]{inputenc}
\usepackage{amsbsy}
\usepackage{amsmath}
\usepackage{amsfonts}
\usepackage{xcolor}
\usepackage{physics}
\usepackage{graphicx}
\begin{document}
\title{The flavor vacuum in the expanding universe and dark matter}

\author{Antonio Capolupo}
\email{capolupo@sa.infn.it}
\affiliation{Dipartimento di Fisica ``E.R. Caianiello'' Universit\`{a} di Salerno, and INFN -- Gruppo Collegato di Salerno, Via Giovanni Paolo II, 132, 84084 Fisciano (SA), Italy}

\author{Sante Carloni}
\email{sante.carloni@unige.it}
\affiliation{DIME Sezione Metodi e Modelli Matematici, Universit\`{a} di Genova,
Via All’Opera Pia 15, 16145 Genova, Italy}

\author{Aniello Quaranta}
\email{anquaranta@unisa.it}
\affiliation{Dipartimento di Fisica ``E.R. Caianiello'' Universit\`{a} di Salerno, and INFN -- Gruppo Collegato di Salerno, Via Giovanni Paolo II, 132, 84084 Fisciano (SA), Italy}

\begin{abstract}
We analyze fermion mixing in the framework of field quantization in curved spacetime. We compute the expectation value of the energy momentum tensor of mixed fermions on the flavor vacuum. We consider spatially flat Friedmann-Lemaitre-Robertson-Walker metrics, and we show that the energy--momentum tensor of the flavor vacuum is diagonal and conserved. Therefore it can be interpreted as the effective energy--momentum tensor of a perfect fluid. In particular, assuming a fixed De Sitter background, the equation of state of the fluid is consistent with that of dust and cold dark matter. Our results establish a new link between quantum effects and classical fluids, and indicate that the flavor vacuum of mixed fermions may represent a new component of dark matter.

\end{abstract}

\maketitle

\section{Introduction}
%

Among the open issues of modern cosmology is the understanding of the dark components of the universe. Most of the total energy density is shared between ``dark energy'', that drives the accelerated expansion of the universe \cite{CMBR}-\cite{SNeIa}, and ``dark matter'', the non--baryonic matter which is responsible for holding galaxies and clusters together \cite{Dm1}-\cite{Dm18}. A conclusive explanation for this ``dark universe'' is still lacking. Several proposals, often very different in nature and perspective, have been put forward to account for dark matter. New particles such as axions, axionlike particles \cite{Axion1,Axion2,Axion01,Axion02,Axion3,Axion03,Axion4,Axion04,Axion5,Axion6,Axion7} and supersymmetric partners \cite{SSDM}, arising from extensions of the standard model, might explain the missing matter. Another possible dark matter component is represented by massive compact objects like primordial black holes \cite{PBH}. 
Other proposals rely on pure quantum field theoretical effects   and the non--trivial structure of the vacuum of flavor fermion mixing \cite{CapDark1,CapDark2,CapDark3,CapDark4,CapDark5}. The relevance of  fermion fields in astrophysical and cosmological contexts is well represented by the role of neutrinos.  These particles are a valuable source of information as they are expected to play an important role in investigating astrophysical processes. They can be used as a test of the standard cosmological model \cite{CosmologicalNeutrinos,PTOLEMY} even at early time \cite{Buchmuller,CosmologicalNeutrinos} and have been recognized as a possible component of dark energy \cite{Mavans}. Moreover the corresponding vacuum has been identified as a possible dark matter component in flat space \cite{CapDark1,CapDark2}.  Neutrinos  are known to oscillate among their three flavors and their general oscillation formulas in curved space-time have been derived in refs. \cite{2020PhRvD.101i5022C,Buoninfante,Grossman,Piriz,Cardall}  together with a number of other  aspects of this phenomenon.

In the following, starting from the analysis of the possible contribution of the flavor vacuum to the dark matter, previously studied in Minkowski spacetime \cite{CapDark1,CapDark2}, we study the behaviour of the flavor vacuum in the case of curved background. We compute, in the context of an homogeneous and isotropic spacetime, represented by a spatially flat Friedmann--Lemaitre--Robertson--Walker metric, the expectation value of the energy momentum tensor of neutrinos on the flavor (mixed) vacuum. We show that this expectation value is a diagonal tensor and satisfies the Bianchi identities. Consequently it behaves as an effective stress energy tensor akin to that of a perfect fluid and can enter the Einstein equations as a regular source term. 
In particular, assuming a fixed De Sitter background, the equation of state of such a fluid is that of dust or cold dark matter ($w=0$). 

These results are not a mere generalization of the studies conducted in flat space \cite{CapDark1,CapDark2,CapDark3,CapDark4,CapDark5}, since the underlying quantum field theory of fermion mixing is much more involved, with respect to flat space, and the properties of the flavor vacuum depend critically on the curved background considered.
The computation performed on the De Sitter background already gives an indication that the flavor vacuum may play a role within the dark sector of the universe. Indeed the flavor vacuum brings along an additional energy density, and, being pressure--less, may be associated with a dark matter component.
According to our results, a possible constituent of dark matter is then represented by the energy of the vacuum of mixed fields. In future works we will analyze the energy--momentum tensor associated to the fermion flavor vacuum in spacetimes that describe large scale structures, like spiral galaxies. In that context the fluid related to the flavor vacuum can be more properly identified with a dark matter component.


The paper is organized as following. Section II contains a summary of the properties of the quantum Dirac equation in curved spacetime.  Section III contains the field quantization of the flavor fields and the introduction of the oscillation components. In Section IV the component of the expectation value of the quantum stress energy tensor is derived. Section V contains the application to a specific background metric i.e.  the de Sitter spacetime and the derivation of  the corresponding exact solution for the component of the stress energy tensor, together with some consideration on its regularization. Finally Section VI is dedicated to the conclusions. 
In the following, we will use the $+---$ signature, and we will assume that greek indices run from $0,...,3$ lower case Latin indices from $1,...,3$ and upper case Latin indices from $1,...,4$. These last indices will represent tetrad indices. The symbol $[,]$ denotes the matrix commutator so that, e.g., $ [\gamma_A,\gamma_B]=\gamma_A\gamma_B-\gamma_B\gamma_A$.

\section{The Dirac Equation and its solutions in flat FLRW spacetime}
For the reader's convenience we start this section by setting the notation and introducing the metric of interest. We then elaborate on the Dirac equation and discuss the general form of the solution.
We will focus on the spatially flat Friedmann-Lema\^{\i}tre-Robertson-Walker spacetime (we use the $+---$ signature) described by the metric
$
 g_{\mu \nu} = \mathrm{diag}\left(1,-C^2(t),-C^2(t),-C^2(t)\right).
$
where $C(t)$ is the scale factor. For our purposes it will be useful to express this metric in terms of the conformal time $\tau$ defined as
$
d\tau = \frac{dt}{C(t)}
$
with range $\tau \in(-\infty,\infty)$ corresponding to $t\in(-\infty,\infty)$.
Using $\tau$, the line element reads
\begin{equation}\label{LineElement}
  ds^2 = C^2(\tau) [d\tau^2 - dx^2 -dy^2 -dz^2]
\end{equation}
when expressed in cartesian spatial coordinates and in conformal time $\tau$. 

In order to write the Dirac equation, we need to choose a tetrad field $e^A_{\mu}$ with the property
$
g_{\mu \nu} = e^A_{\mu}e^{B}_{\nu}\eta_{AB}
$
Given the metric of Eq.(\ref{LineElement}) a convenient choice of tetrads is
\begin{equation}
  e^{A}_{\mu} = C(\tau) \delta^A_{\mu} \ ,
\end{equation}
where the Kronecker symbol $\delta^A_{\mu}$ signals that the only non-zero component of $e^{A}$ is the one for which the Lorentz index $A$ and the spacetime index $\mu$ coincide.

Using the tetrads above one can define the generalization of gamma matrices $\gamma^A$ to the case of curved spacetime 
$
  \tilde{\gamma}^{\mu} = e^{\mu}_A \gamma^A
$

Finally, in order to write the Dirac equation, we also need the spin connections
$
 \omega_{\mu}^{AB} = e^A_{\nu}\Gamma^{\nu}_{\sigma \mu}e^{\sigma B} + e^{A}_{\nu}\partial_{\mu}e^{\nu B} \ .
$
The spinorial covariant derivative is defined with the aid of the spin connections as
$
  D_{\mu} \psi = \partial_{\mu} \psi + \Gamma_{\mu} \psi \,,\ \ \ \ \ \ \ \Gamma_{\mu} = \frac{1}{8} \omega_{\mu}^{AB} [\gamma_A,\gamma_B].
$
We can now write down the Dirac equation
\begin{equation}\label{DiracEquation}
  i \tilde{\gamma}^{\mu}(x) D_{\mu} \psi - m \psi = 0 \ .
\end{equation}
The above equation can be generated by the Lagrangian 
\begin{equation}\label{DiracLagrangian}
  \mathcal{L} = \sqrt{-g} \left\lbrace \frac{i}{2}\left[\bar{\psi} \tilde{\gamma}^{\mu}(x) D_{\mu}\psi - D_{\mu}\bar{\psi} \tilde{\gamma}^{\mu}(x) \psi \right] - m \bar{\psi} \psi \right\rbrace
\end{equation}
and we can define the energy-momentum tensor of the spinor field as the variation of the above Lagrangian with respect to $\psi$ and $\bar{\psi}$ 
\begin{equation}\label{EnergyPulseTensor1}
 T_{\mu \nu} = \frac{i}{2} \left\lbrace \bar{\psi} \tilde{\gamma}_{\mu}(x) D_{\nu}\psi + \bar{\psi} \tilde{\gamma}_{\nu}(x) D_{\mu}\psi - D_{\mu}\bar{\psi} \tilde{\gamma}_{\nu}(x) \psi - D_{\nu}\bar{\psi} \tilde{\gamma}_{\mu}(x) \psi\right\rbrace \ .
\end{equation}
In the metric \eqref{LineElement} the Dirac equation reads
\begin{equation}\label{DiracEquation2}
  \left(i \gamma^0 \partial_{\tau} + \frac{3i}{2}\frac{\partial_{\tau}C}{C}\gamma^0 + i \gamma^j \partial_j - mC \right)\psi = 0 \ .
\end{equation}
We remark that the Dirac equation \eqref{DiracEquation2} holds for the quantum field theoretic free Dirac field of mass $m$. At odds with the \emph{classical} Dirac field, we cannot drop the spatial derivatives by arguing that space-dependent quantities cannot enter the right hand side of the Einstein equation if the metric depends only on time. Indeed \emph{all} the solutions to Eq. \eqref{DiracEquation2} must be considered for the field expansion, including those which depend explicitly on the spatial coordinates. Consistency with the (time-dependent only) metric is then achieved for the expectation value of the physical observables, including the energy-momentum tensor, which turn out to depend only upon time.   

The spatial dependence of this equation suggests that we look for plane wave solutions $\psi \propto e^{i \pmb{p} \cdot \pmb{x}}$, where the mode label $\pmb{p}$ is a $3$-vector that can be thought as the would-be plane wave momentum when $C(\tau)$ reduces to a constant, and "$\pmb{a} \cdot \pmb{b}$" is a shorthand notation for $\sum_{j=1,2,3} a_j b_j $. We remark that the actual momentum that is istanteneously carried by the mode with label $\pmb{p}$ is the comoving momentum\footnote{This is most conveniently seen by inserting the plane wave ansatz in Eq. \eqref{DiracEquation2} and reverting to the coordinate time $t$:
\begin{equation*}
\left[ i \gamma^0 \left(\partial_t + \frac{3 i}{2} \frac{\partial_t C}{C} \right) - \frac{\gamma^j p_j}{C} - m \right] \psi_{\pmb{p}} = 0 \ .
\end{equation*}
In this form the equation resembles the Dirac equation in flat space, with a time-dependent potential $\frac{3 i}{2} \frac{\partial_t C}{C}$ and with the quantity $\frac{p_j}{C}$ playing the role of an instantaneous momentum.} $\frac{\pmb{p}}{C(\tau)}$.

Using the helicity eigenbispinors $\xi_{\lambda} $ defined as 
\begin{equation}
 \frac{\pmb{\sigma}\cdot \pmb{p}}{p}\xi_{\lambda} = \lambda \xi_{\lambda}\ ,
\end{equation}
the solution of \eqref{DiracEquation2} can be written as the combination of the positive and negative energy solutions in the form
\begin{equation}\label{Solutions}
 u_{\pmb{p},\lambda}(\tau,\pmb{x}) = e^{i \pmb{p} \cdot \pmb{x}} \begin{pmatrix} f_{p}(\tau) \xi_{\lambda} (\hat{p}) \\
  g_{p}(\tau) \lambda \xi_{\lambda} (\hat{p})
  \end{pmatrix}  \ \ \ \ v_{\pmb{p},\lambda}(\tau,\pmb{x})= e^{i \pmb{p} \cdot \pmb{x}} \begin{pmatrix} g^*_{p}(\tau) \xi_{\lambda} (\hat{p}) \\
  -f^*_{p}(\tau) \lambda \xi_{\lambda} (\hat{p})
  \end{pmatrix}    \ .
\end{equation}
Here $\hat{p} = \frac{\pmb{p}}{p}$ denotes the unit vector in the direction of $\pmb{p}$, $\pmb{\sigma}$ is the vector of Pauli matrices and $\lambda = \pm 1$. Notice that for each $\lambda=\pm 1$, the quantity $\xi^{\dagger}_{\lambda}(\hat{p})\sigma_i \xi_{\lambda}(\hat{p})$ is an odd function of $p_i$, for $i=1,2,3$. In particular, $\xi^{\dagger}_{\lambda}(\hat{p})\sigma_i \xi_{\lambda}(\hat{p})$ changes sign when the momentum is reversed $\pmb{p}\rightarrow - \pmb{p}$.
 The statement can be proven by direct calculation, as shown in Appendix B. 
Inserting Eq.(\ref{Solutions}) in Eq.(\ref{DiracEquation2}), and using the defining property of the helicity bispinors, we can write \eqref{DiracEquation2} as the system 
\begin{equation}\label{DiracEquation3}
\begin{split}
i \partial_{\tau}f_{\pmb{p}} & = \left(mC -\frac{3i}{2}\frac{\partial_{\tau}C}{C} \right)f_{\pmb{p}} +pg_{\pmb{p}} 
\\
 i \partial_{\tau}g_{\pmb{p}} & = \left(-mC -\frac{3i}{2}\frac{\partial_{\tau}C}{C} \right)g_{\pmb{p}} +pf_{\pmb{p}} \ , \\
i \partial_{\tau}f^*_{p} & = \left(-mC -\frac{3i}{2}\frac{\partial_{\tau}C}{C} \right)f^*_{p} -pg^*_{p} \\ 
 i \partial_{\tau}g^*_{p} & = \left(mC -\frac{3i}{2}\frac{\partial_{\tau}C}{C} \right)g^*_{p} -pf^*_{p} \ .
 \end{split}
\end{equation}
The above equations show that the functions $f,g$ depend only on the modulus $p$ of the $3$-momentum. Thus, from now on, we will drop the vector index $\pmb{p}$ in favor of the scalar index $p$.

The scalar product between two solutions of the Dirac equation $A, B$ is defined as
\begin{equation}\label{ScalarProduct}
  (A,B)_{\tau}=\int_{\Sigma_{\tau}}d^3x \sqrt{-g}\bar{A}\tilde{\gamma}^{\tau} (x) B \ ,
\end{equation}
where the integration is to be carried over a hypersurface $\Sigma_{\tau}$ of constant conformal time $\tau$. If $A$ and $B$ are solutions of the \textit{same} Dirac equation, the scalar product $(A,B)_{\tau}$ is independent of $\tau$. This is no longer true if $A$ and $B$ are solutions of distinct Dirac equations. Taking into account that the determinant is $g =- C^8 $ and adopting the normalization
\begin{equation}\label{Normalization1}
  |f_p|^2+|g_p|^2 =\frac{1}{\left(2\pi C\right)^3}
\end{equation}
 we obtain the following orthonormality and completeness relations, respectively (for details, see Appendix A):
\begin{eqnarray}\label{ScalarProduct2}
   (u_{\pmb{p},\lambda},u_{\pmb{q},\lambda'})_{\tau} = \delta^{3}(\pmb{p}-\pmb{q})\delta_{\lambda,\lambda'} \ ,
 \qquad
 (u_{\pmb{p},\lambda},v_{\pmb{q},\lambda'})_{\tau} = (v_{\pmb{p},\lambda},u_{\pmb{q},\lambda'})_{\tau}=0 \ ,
 \qquad
  (v_{\pmb{p},\lambda},v_{\pmb{q},\lambda'})_{\tau} = \delta^{3}(\pmb{p}-\pmb{q})\delta_{\lambda,\lambda'} \ ,
  \end{eqnarray}
\begin{eqnarray}\label{Completeness1}
\sum_{\lambda} \left(u_{\pmb{p},\lambda}u^{\dagger}_{\pmb{p},\lambda} + v_{\pmb{p},\lambda}v^{\dagger}_{\pmb{p},\lambda}\right) &=& \frac{1}{\left(2\pi C\right)^3} \begin{pmatrix} \mathbb{I} & 0 \\ 0 & \mathbb{I} \end{pmatrix} \ .
\end{eqnarray}
 Eq.(\ref{Completeness1}) shows that the set of solutions (\ref{Solutions}) provides a basis not only for the solution space of the Dirac equation, but also for the space of $4$-spinors.

The system of equations (\ref{DiracEquation3}) can be further simplified by introducing the functions
\begin{equation}
  \phi_p  = C^{\frac{3}{2}} f_p \ \ \ \ \gamma_p  = C^{\frac{3}{2}} g_p 
\end{equation}
since then the system becomes
\begin{equation}\label{DiracEquation5}
\begin{split}
\nonumber \partial_{\tau}\phi_p &= -imC \phi_p - ip\gamma_p \\ 
\partial_{\tau}\gamma_p &= imC \gamma_p - ip\phi_p \\
\nonumber \partial_{\tau}\phi^*_p &= imC \phi^*_p + ip\gamma^*_p \\ 
\partial_{\tau}\gamma^*_p &= -imC \gamma^*_p + ip\phi^*_p 
\end{split}
\end{equation}
and the normalization condition (\ref{Normalization1}) becomes
\begin{equation}\label{Normalization2}
 |\phi_p|^2 + |\gamma_p|^2 = \frac{1}{(2\pi)^3} \ .
\end{equation}
The first two of Eqs.(\ref{DiracEquation5}) can be combined to give two second order equation for $\phi_p$ 
\begin{equation}\label{PhiEquation}
\partial_{\tau}^2 \phi_p + \left(im\partial_{\tau}C + p^2 + m^2 C^2 \right)\phi_p = 0 \ .
\end{equation}
In the same way, we can obtain a second-order equation for $\phi^*_p$.

We conclude the section by introducing a bilinear form of the solutions of the Dirac Equation, which is convenient
for the study of the energy-momentum tensor. Given two solutions $A, B$ of the Dirac equation, we define the auxiliary tensor as
\begin{equation}{\label{AuxiliaryTensor1}}
  L_{\mu \nu}(A,B)=\bar{A} \tilde{\gamma}_{\mu}(x) D_{\nu}B + \bar{A} \tilde{\gamma}_{\nu}(x) D_{\mu}B - D_{\mu}\bar{A} \tilde{\gamma}_{\nu}(x) B - D_{\nu}\bar{A} \tilde{\gamma}_{\mu}(x) B \ .
\end{equation}
The properties of the auxiliary tensor are analyzed in Appendix C.

\section{Quantization of flavor fields}

In the following, we start by quantizing a single Dirac field of definite mass, and then we report the analysis of the main properties of two flavor mixed fields. For more details, we refer to \cite{2020PhRvD.101i5022C}.

\subsection{Dirac field }
Since the set of solutions (\ref{Solutions}) $\lbrace u_{\pmb{p},\lambda},v_{\pmb{p},\lambda} \rbrace_{\pmb{p},\lambda}$ is complete, any solution of the (linear) Dirac equation can be written as a linear combination of these modes. In particular, this is true for the Dirac field:
$\label{FieldExpansion}
  \psi (x) = \sum_{\lambda} \int d^3 p \left(A_{\pmb{p,\lambda}} u_{\pmb{p},\lambda} + B^{*}_{-\pmb{p},\lambda} v_{\pmb{p},\lambda} \right)  
$.
In this equation, the notation $B^*_{-\pmb{p},\lambda}$ is chosen to remark that $v_{\pmb{p},\lambda}$ describes an antiparticle with momentum $-\pmb{p}$. Notice that all the spacetime dependence is in the modes, while the coefficients are independent of both space and time coordinates.

Quantization is achieved, as usual, by promoting the field, and thus the expansion coefficients to operators
\begin{equation}\label{FieldExpansion2}
  \psi (x) = \sum_{\lambda} \int d^3 p \left(A_{\pmb{p,\lambda}} u_{\pmb{p},\lambda} + B^{\dagger}_{-\pmb{p},\lambda} v_{\pmb{p},\lambda} \right) \ .
\end{equation}
and imposing the canonical anticommutation relations. The momentum conjugate to $\psi(x)$, according to the Lagrangian (\ref{DiracLagrangian}), is
$
  \pi_{\psi} (x) =i C^3 \psi^{\dagger} (x) \ ,
$
so that the canonical anticommutation relations to be imposed are
\begin{equation}\label{AnticommutationRelations}
  \lbrace \psi_{A}(\tau,\pmb{x}) , \pi_{\psi B} (\tau,\pmb{x'}) \rbrace = iC^3 \lbrace \psi_{A}(\tau,\pmb{x}) , \psi^{\dagger}_{B} (\tau,\pmb{x'}) \rbrace = i\delta_{AB}\delta^3 (\pmb{x}-\pmb{x'}) \ .
\end{equation}
Here the indices $A,B$ are referred to the spinor components $A,B=1,2,3,4$. It is easy to show that the relations (\ref{AnticommutationRelations}) are satisfied if one imposes the following anticommutation relations on the coefficients:
\begin{equation}
  \lbrace A_{\pmb{p,\lambda}}, A^{\dagger}_{\pmb{q},\lambda'} \rbrace =  \lbrace B_{\pmb{p,\lambda}}, B^{\dagger}_{\pmb{q},\lambda'} \rbrace = \delta_{\lambda\lambda'} \delta^3 (\pmb{p}-\pmb{q})
\end{equation}
with all the other anticommutators vanishing (see Appendix ...). The field expansion (\ref{FieldExpansion2}) defines the vacuum state $\ket{0}$ as the state annihilated by all the annihilation operators
$
  A_{\pmb{p},\lambda} \ket{0} = B_{\pmb{p},\lambda} \ket{0}= 0 \,, \ \ \ \ \forall \pmb{p},\lambda \ .
$
It is important to stress that the definition of the vacuum state depends critically on the choice of the field modes. Specifically, its particle interpretation is tied to the boundary conditions specified on the solutions (\ref{Solutions}). Another kind of field expansion is possible if one assumes a specific time evolution of the modes, as is done within the adiabatic approximation (see, e.g., \cite{Parker69, PostParker1, PostParker2, PostParker3}). Contrary to our field expansion (\ref{FieldExpansion2}), the annihilation operators (and thus the vacuum) are thereby endowed with a specific time dependence. The expansion (\ref{FieldExpansion2}), instead, does not assume any particular time dependence and is, therefore, more general. The two expansions can be made to coincide at a given time.
The quantized energy-momentum tensor is obtained by inserting the field expansion in the definition (\ref{EnergyPulseTensor1}):
\begin{eqnarray}\label{EnergyPulseTensor2}
 \nonumber  T_{\mu \nu} 
&=& \frac{i}{2}\sum_{\lambda,\lambda'} \int d^3 p \int d^3 q \lbrace A^{\dagger}_{\pmb{p},\lambda}A_{\pmb{q},\lambda'} L_{\mu \nu} (u_{\pmb{p},\lambda},u_{\pmb{q},\lambda'}) + A^{\dagger}_{\pmb{p},\lambda}B^{\dagger}_{-\pmb{q},\lambda'} L_{\mu \nu} (u_{\pmb{p},\lambda},v_{\pmb{q},\lambda'}) \\
&& + B_{-\pmb{p},\lambda}A_{\pmb{q},\lambda'} L_{\mu \nu} (v_{\pmb{p},\lambda},u_{\pmb{q},\lambda'}) + B_{-\pmb{p},\lambda}B^{\dagger}_{-\pmb{q},\lambda'} L_{\mu \nu} (v_{\pmb{p},\lambda},v_{\pmb{q},\lambda'})\rbrace \ ,
\end{eqnarray}
where we have used the definition (\ref{AuxiliaryTensor1}) of $L_{\mu \nu}$. The Hamiltonian density, defined with respect to $\partial_{\tau}$ corresponds to the component $T_{\tau}^{\tau} $ of the above equation. It follows immediately that the vacuum is not an eigenstate of the Hamiltonian unless $L^{\tau}_{\tau}(v_{\pmb{p},\lambda},u_{\pmb{q},\lambda'}) = 0$, due to the $A^{\dagger}_{\pmb{p},\lambda}B^{\dagger}_{-\pmb{q},\lambda'}$ term. In particular, the vacuum is unstable under the creation of particle-antiparticle pairs with opposite momentum, a result known from the previous analyses. This is a reflection of the non-invariance under time translations. On the other hand, from the relation: $
L_{\tau i}(u_{\pmb{p},\lambda},v_{\pmb{p},\lambda}) = 0
$, derived in Appendix C (cfr.  Eq.(\ref{ZeroMomentum})), it is clear that the vacuum respects the residual translational symmetry in the spatial coordinates, i.e. $\pmb{P}\ket{0} = 0$.

\subsection{The flavor fields}
Up to now, our considerations have been restricted to a single Dirac field of definite mass. To introduce the flavor fields, we follow \cite{2020PhRvD.101i5022C} and start by introducing two Dirac fields with distinct masses $m_1,m_2$. We consider two flavors for simplicity but the analysis can be easily generalized to three flavors. The theory of two free massive Dirac fields is just the product of two copies of the theory for a single Dirac field. All the relations discussed above remain valid, provided we assign a new index $j=1,2$ to all the quantities involved $\psi_j,m_j,A_{\pmb{p},\lambda;j}, u_{\pmb{p},\lambda;j}$ and so on. All the previous relations are index-wise valid for $j=1,2$. The mass vacuum that we now denote explicitly as $\ket{0_M}$ is annihilated by all the annihilators for each index $j$. The total energy-momentum tensor is simply the sum of two copies of Eq.(\ref{EnergyPulseTensor2}), one for each $j$.
We also require that each mode of the field $1$ is related to the corresponding mode of the field $2$, with the same labels $\pmb{p},\lambda$ by the substitution $m_1 \rightarrow m_2$, and vice-versa.

The flavor fields are then introduced via the rotation
\begin{eqnarray}\nonumber
  \psi_{e}(x) &=& \cos(\theta) \psi_{1}(x) + \sin(\theta) \psi_2(x)
  \\
  \ \psi_{\mu}(x) &=& \cos(\theta) \psi_{2}(x) - \sin(\theta) \psi_1(x) \ ,
\end{eqnarray}
where $\theta$ is the two-flavor mixing angle. At the quantum level, the rotation is employed by the mixing generator
\begin{equation}\label{Generator}
  \mathcal{I}_{\theta} (\tau)= exp\left \lbrace\theta \left[ (\psi_1,\psi_2)_{\tau} - (\psi_2,\psi_1)_{\tau} \right] \right \rbrace \ .
\end{equation}
Here $(\psi_2,\psi_1)_{\tau}$ stands for the scalar product at the $\tau$ hypersurface (recall that for fields of distinct masses the product \emph{does} depend on $\tau$). The flavor fields are then
\begin{eqnarray}\nonumber
  \psi_{e}(x) &=& \mathcal{I}^{-1}_{\theta}  \psi_1 (x) \mathcal{I}_{\theta} 
   \\
   \psi_{\mu}(x) &=& \mathcal{I}^{-1}_{\theta}  \psi_2 (x) \mathcal{I}_{\theta} \ .
\end{eqnarray}
The action of the generator also defines the flavor annihilators ($A_{\pmb{p},\lambda;e}  = \mathcal{I}^{-1}_{\theta}  A_{\pmb{p},\lambda;1} \mathcal{I}_{\theta} $ and similar) and the flavor vacuum, as
\begin{equation}
  \ket{0_F } = \mathcal{I}^{-1}_{\theta}  \ket{0_M} \ .
\end{equation}
Notice that, contrary to the mass vacuum, the flavor vacuum has an explicit $\tau$ dependence. The terminology 'flavor vacuum' is justified in that this state is annihilated by all the flavor annihilators.

\section{VEV of the energy-momentum tensor on the flavor vacuum}

We are specifically interested in the contributions that flavor mixing induces on the energy-momentum tensor. More precisely, we ask what is the expectation value of the energy-momentum tensor on the state corresponding to the absence of flavor neutrinos at a given time $\ket{O_F(\tau_0)}$. We stress that in this calculation, we assume a fixed but arbitrary expansion of the mass fields, and therefore a fixed but arbitrary choice of the mass vacuum. The effect of a change in the mass representation on the flavor fields is known \cite{2020PhRvD.101i5022C}, and once the result is computed for a given representation, one can implement the adequate transformations to get the result in other mass representations. Likewise, we keep the time $\tau_0$ arbitrary and distinct from the time argument $\tau$ of the energy-momentum tensor. The quantity we wish to compute is then

\begin{equation}\label{VeV}
 \mathbb{T}_{\mu \nu} = \bra{0_F(\tau_0)} T_{\mu \nu} \ket{0_F(\tau_0)} \ ,
\end{equation}

where $T_{\mu \nu}$ is given by Eq.(\ref{EnergyPulseTensor2}).
We remark that the only sensible definition for the energy-momentum tensor is the one (\ref{EnergyPulseTensor2}) in terms of the fields with definite mass. Let us analyze the typical term in equation (\ref{VeV}). It has the form
\begin{equation}
  \bra{0_F(\tau_0)} A^{\dagger}_{\pmb{p},\lambda;1} A_{\pmb{q},\lambda';1} \ket{0_F(\tau_0)} \ .
\end{equation}
Using the definition of the flavor vacuum this equals
\begin{eqnarray}
 \nonumber \bra{0_M}\mathcal{I}_{\theta}(\tau_0) A^{\dagger}_{\pmb{p},\lambda;1} A_{\pmb{q},\lambda';1}\mathcal{I}^{-1}_{\theta}(\tau_0) \ket{0_M} &= &  \nonumber  \bra{0_M}\mathcal{I}_{\theta}(\tau_0) A^{\dagger}_{\pmb{p},\lambda;1} \mathcal{I}^{-1}_{\theta}(\tau_0) \mathcal{I}_{\theta}(\tau_0) A_{\pmb{q},\lambda';1}\mathcal{I}^{-1}_{\theta}(\tau_0) \ket{0_M}
 \\ &=&
\bra{0_M}\mathcal{I}^{-1}_{-\theta}(\tau_0) A^{\dagger}_{\pmb{p},\lambda;1} \mathcal{I}_{-\theta}(\tau_0) \mathcal{I}^{-1}_{-\theta}(\tau_0) A_{\pmb{q},\lambda';1}\mathcal{I}_{-\theta}(\tau_0) \ket{0_M}
\end{eqnarray}
where we have used that (eq. \ref{Generator}) $\mathcal{I}^{-1}_{\theta} = \mathcal{I}_{-\theta} $. Now the operator $\mathcal{I}^{-1}_{-\theta}(\tau_0) A_{\pmb{p},\lambda;1} \mathcal{I}_{-\theta}(\tau_0)$ is just the mass annihilator transformed according to a mixing transformation with angle $-\theta$. Knowing the transformation rule in terms of $\theta$ \cite{2020PhRvD.101i5022C} we can easily write down the transformed operators:

\begin{equation}\label{MixingTransformations}
\begin{split}
\mathcal{I}_{-\theta}^{-1} (\tau_0) A_{\pmb{p},\lambda;1} \mathcal{I}_{-\theta}(\tau_0) &= \cos(\theta)A_{\pmb{p},\lambda;1}-\sin(\theta) \left(\Lambda^*_{\pmb{p}}(\tau_0) A_{\pmb{p},\lambda;2} + \Xi_{\pmb{p}}(\tau_0)B^{\dagger}_{-\pmb{p},\lambda;2} \right) \\
\mathcal{I}_{-\theta}^{-1} (\tau_0) A_{\pmb{p},\lambda;2} \mathcal{I}_{-\theta}(\tau_0) &= \cos(\theta)A_{\pmb{p},\lambda;2}+\sin(\theta) \left(\Lambda_{\pmb{p}}(\tau_0) A_{\pmb{p},\lambda;1} - \Xi_{\pmb{p}}(\tau_0)B^{\dagger}_{-\pmb{p},\lambda;1} \right) \\
 \mathcal{I}_{-\theta}^{-1} (\tau_0) B_{-\pmb{p},\lambda;1} \mathcal{I}_{-\theta}(\tau_0) &= \cos(\theta)B_{-\pmb{p},\lambda;1}-\sin(\theta) \left(\Lambda^*_{\pmb{p}}(\tau_0) B_{-\pmb{p},\lambda;2} - \Xi_{\pmb{p}}(\tau_0)A^{\dagger}_{\pmb{p},\lambda;2} \right) \\
 \mathcal{I}_{-\theta}^{-1} (\tau_0) B_{-\pmb{p},\lambda;2} \mathcal{I}_{-\theta}(\tau_0) &= \cos(\theta)B_{-\pmb{p},\lambda;2}+\sin(\theta) \left(\Lambda_{\pmb{p}}(\tau_0) B_{-\pmb{p},\lambda;1} + \Xi_{\pmb{p}}(\tau_0)A^{\dagger}_{\pmb{p},\lambda;1} \right)  
\end{split}
\end{equation}
while the transformation rule for the adjoint operators can be obtained by considering the adjoint equations. The Bogoliubov coefficients are defined as the inner products
\begin{equation}\label{BogoliubovCoefficients}
\begin{split}
 \delta^3 (\pmb{0})\Lambda_{\pmb{p}}(\tau) &= \left(u_{\pmb{p},\lambda;2},u_{\pmb{p},\lambda;1}\right)_{\tau} = \left(v_{\pmb{p},\lambda;1},v_{\pmb{p},\lambda;2}\right)_{\tau} \\
  \delta^3 (\pmb{0})\Xi_{\pmb{p}}(\tau) &= \left(u_{\pmb{p},\lambda;1},v_{\pmb{p},\lambda;2}\right)_{\tau} = - \left(u_{\pmb{p},\lambda;2},v_{\pmb{p},\lambda;1}\right)_{\tau} \ ,
  \end{split}
\end{equation}
where the notation anticipates that they don't depend on $\lambda$ (we will see that they actually depend only on the modulus of the momentum $p$). The $\delta^3(\pmb{0})$ factor is a reminiscence of the more general expressions involving distinct momenta $\Lambda_{\pmb{p},\pmb{q}} \propto \delta^{3}(\pmb{p}-\pmb{q})$ \cite{2020PhRvD.101i5022C}. The delta factor is absorbed by a corresponding momentum integration in the Eqs.(\ref{MixingTransformations}), leaving only the finite coefficients defined via Eq.(\ref{BogoliubovCoefficients}). These coefficients satisfy $|\Lambda_{\pmb{p}} |^2 + |\Xi_{\pmb{p}} |^2=1$ for all $\pmb{p},\tau$. With the aid of Eqs.(\ref{MixingTransformations}), we can evaluate all the expectation values appearing in Eq.(\ref{VeV}):
\begin{equation}
\begin{split}
  \bra{0_F(\tau_0)}A^{\dagger}_{\pmb{p},\lambda;j}A_{\pmb{q};\lambda';j}\ket{0_F(\tau_0)} &= \sin^2 \theta |\Xi_{\pmb{p}}(\tau_0)|^2 \delta_{\lambda \lambda'} \delta^3 (\pmb{p}-\pmb{q})\ \,, \ \ \ \forall j \\
   \bra{0_F(\tau_0)}B^{\dagger}_{-\pmb{p},\lambda;j}B_{-\pmb{q};\lambda';j}\ket{0_F(\tau_0)} &= \sin^2 \theta |\Xi_{\pmb{p}}(\tau_0)|^2 \delta_{\lambda \lambda'} \delta^3 (\pmb{p}-\pmb{q})\ \,, \ \ \ \forall j \\
 \bra{0_F(\tau_0)}A^{\dagger}_{\pmb{p},\lambda;1}B^{\dagger}_{-\pmb{q};\lambda';1}\ket{0_F(\tau_0)} &=\sin^2 \theta \Xi^*_{\pmb{p}}(\tau_0) \Lambda_{\pmb{p}}(\tau_0) \delta_{\lambda \lambda'} \delta^3 (\pmb{p}-\pmb{q}) \\
 \bra{0_F(\tau_0)}A^{\dagger}_{\pmb{p},\lambda;2}B^{\dagger}_{-\pmb{q};\lambda';2}\ket{0_F(\tau_0)} &= -\sin^2 \theta \Xi^*_{\pmb{p}}(\tau_0) \Lambda^*_{\pmb{p}}(\tau_0) \delta_{\lambda \lambda'} \delta^3 (\pmb{p}-\pmb{q}) \\
 \bra{0_F(\tau_0)}B_{-\pmb{p},\lambda;1}A_{\pmb{q};\lambda';1}\ket{0_F(\tau_0)} &= \sin^2 \theta \Xi_{\pmb{p}}(\tau_0) \Lambda^*_{\pmb{p}}(\tau_0) \delta_{\lambda \lambda'} \delta^3 (\pmb{p}-\pmb{q}) \\
 \bra{0_F(\tau_0)}B_{-\pmb{p},\lambda;2}A_{\pmb{q};\lambda';2}\ket{0_F(\tau_0)} &= -\sin^2 \theta \Xi_{\pmb{p}}(\tau_0) \Lambda_{\pmb{p}}(\tau_0) \delta_{\lambda \lambda'} \delta^3 (\pmb{p}-\pmb{q}) \ . \\
\end{split}
\end{equation}
It is convenient to give the result by splitting out the pure mixing component:
\begin{eqnarray}\label{VeV2}
\mathbb{T}_{\mu \nu} & =& \mathbb{T}_{\mu \nu}^{(MIX)} + \mathbb{T}_{\mu \nu}^{(N)} \\
\nonumber\mathbb{T}_{\mu \nu}^{(MIX)} &=& \frac{i}{2} \sin^2 \theta \,\sum_{\lambda}\int d^3 p \lbrace |\Xi_{\pmb{p}}(\tau_0)|^2 \sum_{j=1,2} \left( L_{\mu \nu}(u_{\pmb{p},\lambda;j},u_{\pmb{p},\lambda;j}) - L_{\mu \nu}(v_{\pmb{p},\lambda;j},v_{\pmb{p},\lambda;j})\right) \\
\nonumber & +& \ \Xi^*_{\pmb{p}}(\tau_0)\Lambda_{\pmb{p}}(\tau_0) L_{\mu \nu}(u_{\pmb{p},\lambda;1},v_{\pmb{p},\lambda;1}) + \Xi_{\pmb{p}}(\tau_0) \Lambda^*_{\pmb{p}}(\tau_0) L_{\mu \nu}(v_{\pmb{p},\lambda;1},u_{\pmb{p},\lambda;1}) \\
 & - &\Xi^*_{\pmb{p}}(\tau_0)\Lambda^*_{\pmb{p}}(\tau_0) L_{\mu \nu}(u_{\pmb{p},\lambda;2},v_{\pmb{p},\lambda;2}) - \Xi_{\pmb{p}}(\tau_0) \Lambda_{\pmb{p}}(\tau_0) L_{\mu \nu}(v_{\pmb{p},\lambda;2},u_{\pmb{p},\lambda;2}) \rbrace \\
\mathbb{T}_{\mu \nu}^{(N)} & =& \frac{i}{2}\sum_{\lambda} \sum_{j=1,2}\int d^3 p L_{\mu \nu}(v_{\pmb{p},\lambda;j},v_{\pmb{p},\lambda;j}) \ .
\end{eqnarray}
The last term comes from applying the anticommutator relation $\left\lbrace B^{\dagger}_{-\pmb{p},\lambda;j}, B_{-\pmb{q},\lambda';j} \right\rbrace = \delta_{\lambda \lambda'} \delta^3(\pmb{p}-\pmb{q})$ to the $BB^{\dagger}$ term. While the remaining terms, regrouped under the symbol $\mathbb{T}_{\mu \nu}^{(MIX)}$ show an explicit dependence on $\sin^2 \theta$, and are therefore zero in absence of mixing, the last term is present independently of mixing. Indeed, it is easy to check that $\mathbb{T}^{(N)}_{\mu \nu}$ represents the expectation value of the energy-momentum tensor on the \emph{mass} vacuum:
\begin{equation}
  \mathbb{T}^{(N)}_{\mu \nu} = \bra{0_M} T_{\mu \nu} \ket{0_M} \ .
\end{equation}
The origin of this term is an ordering ambiguity in the energy-momentum tensor of the quantized fields. To understand its significance, let us consider the flat space limit, where $v_{\pmb{p}, \lambda; j} \propto e^{i \omega_{p;j} t}$ with $\omega_{p;j} = \sqrt{p^2 + m_j^2}$. The auxiliary tensor becomes 
\begin{equation}
 L_{\mu \nu} (A,B) = \bar{A} \gamma_{\mu} \partial_{\nu} B + \bar{A}\gamma_{\nu} \partial_{\mu} B - \partial_{\mu} \bar{A}\gamma_{\nu} B - \partial_{\nu} \bar{A} \gamma_{\mu} B 
\end{equation}
where both the Dirac matrices and the derivatives are the ordinary flat space ones. In particular one has
\begin{equation}
 L_{00} (v_{\pmb{p},\lambda;j},v_{\pmb{p},\lambda;j}) = 2 (v_{\pmb{p},\lambda;j}^{\dagger} \partial_t v_{\pmb{p},\lambda;j} - \partial_t v_{\pmb{p},\lambda;j}^{\dagger}  v_{\pmb{p},\lambda;j}) = 4 i \omega_{p;j} v_{\pmb{p},\lambda;j}^{\dagger}  v_{\pmb{p},\lambda;j} = 4 i \omega_{p;j} \ ,
\end{equation}
where we have made use of the normalization $v_{\pmb{p},\lambda;j}^{\dagger}  v_{\pmb{p},\lambda;j} = 1$. Then 
\begin{equation}
  \mathbb{T}^{(N)}_{0 0} = - 2 \sum_{\lambda} \sum_{j=1,2} \int d^3p \  \omega_{p;j} \ \ \ \ \ \ \ (<0)     \  . 
\end{equation}
We can see that $\mathbb{T}^{(N)}_{0 0}$ corresponds to the diverging negative energy that is removed by the normal ordering prescription in flat space. In order that the energy-momentum tensor $T_{\mu \nu}$ approaches the normal-ordered flat-space expression in the limit, one must subtract  $\mathbb{T}^{(N)}_{\mu \nu}$. Then we define a renormalized energy-momentum tensor by
\begin{equation}
 T_{\mu \nu}^r = T_{\mu \nu} - \mathbb{T}^{(N)}_{\mu \nu}
\end{equation}
and whose expectation value is
\begin{equation}
 \bra{0_F(\tau_0)}T_{\mu \nu}^r\ket{0_F(\tau_0)} = \mathbb{T}_{\mu \nu}^{(MIX)} \ . 
\end{equation}
We now derive the general properties of the Bogoliubov coefficients, we show that $\mathbb{T}^{(MIX)}_{\mu \nu}$ behaves as a perfect fluid and determine the functional expression of $\mathbb{T}_{\tau \tau}$ and $\mathbb{T}^{\mu}_{\mu}$.

\subsection{General properties of the Bogoliubov Coefficients}
Much can still be said without specifying the precise metric, i.e., without providing an explicit form for the scale factor $C $. Plugging the solutions (\ref{Solutions}) in the definition (\ref{BogoliubovCoefficients}) we 
have
\begin{eqnarray}\label{BogoliubovCoefficients2}
\Lambda_p (\tau) &=& (2\pi C)^3 \left(f_{p;2}^*  f_{p;1} + g^{*}_{p;2} g_{p;1} \right) \ ,
\end{eqnarray}
where we have switched to the notation $\Lambda_p$ to highlight that it depends only upon the modulus $p$. In a similar way
\begin{eqnarray}\label{BogoliubovCoefficients3}
\Xi_p (\tau) &=& (2\pi C)^3 \left(f_{p;1}^*  g^*_{p;2} - g^{*}_{p;1} f^*_{p;2} \right) \ .
\end{eqnarray}
From  Eqs.(\ref{BogoliubovCoefficients2}) and (\ref{BogoliubovCoefficients3}) we can verify that \begin{eqnarray}\label{BogoliubovCoefficients4}
|\Lambda_{p}(\tau)|^2 + |\Xi_p (\tau)|^2 &=& 
1 \ ,
\end{eqnarray}
where in the last step we have made use of the normalization condition (\ref{Normalization}).
We conclude this subsection by expressing the Bogoliubov coefficients in terms of the reduced functions $\phi_p,\gamma_p$:
\begin{eqnarray}\label{BogoliubovCoefficients5}
\nonumber \Lambda_p (\tau) &=& (2\pi)^3 \left( \phi_{p;2}^* \phi_{p;1} + \gamma_{p;2}^*  \gamma_{p;1} \right) \\
\Xi_p (\tau) &=& (2\pi)^3 \left( \phi_{p;1}^* \gamma^*_{p;2} - \gamma_{p;1}^*  \phi^*_{p;2} \right) \ .
\end{eqnarray}

\subsection{Diagonality of the energy-momentum tensor}
Using the result (Eq.(\ref{BogoliubovCoefficients5})), we can prove that $\mathbb{T}_{\mu \nu}$ is non-zero only when $\mu=\nu$. This result has the important consequence that $\mathbb{T}^{\mu}_{\nu}$ can be interpreted as the energy-momentum tensor of a perfect fluid. 

We start by proving that $\mathbb{T}_{\tau i} = 0$ for $i=1,2,3$. 
Since we can always write $L_{\tau i} (a,b) = p_i h_{a,b} (p) \ ,$ (see Appendix C for details),
each of the terms in the integrand of Eq. (\ref{VeV2}) is of the form $p_i \mathcal{F}(p)$ , with $\mathcal{F}(p)$ a function of $p$ only. We therefore have an odd function of $p_i$ integrated over an even domain $p_i \in (-\infty,+\infty)$ and the integral vanishes:
\begin{equation}
  \mathbb{T}_{\tau i} = 0 \ .
\end{equation}
The situation is similar for $\mathbb{T}_{ij}$ with $i\neq j$ and $i,j=1,2,3$. In this case (see Eq.(\ref{OffDiagonal3}) of appendix C), namely, $
  L_{ij}(a,b) = p_i p_j l_{a,b}(p)
$, implies that each term under the integral sign is of the form $p_i p_j\mathcal{G}(p)$ with $\mathcal{G}(p)$ a function of $p$ alone. It is clear that also in this case, the integral over the even domain $(p_i,p_j) \in (-\infty,+\infty)\times(-\infty,+\infty)$ vanishes, yielding
$
  \mathbb{T}_{i j} = 0 $ for $  i \neq j 
$.
Notice that the same conclusion does not apply to $\mathbb{T}_{ii}$, since in this case the integrand is an even function $p_i^2 \mathcal{A} (p)$ of $p_i$. It is easy to verify that $\mathbb{T}_{ii}$ is the same for each $i = 1,2,3$,consistently with the isotropy of the underlying metric.
Due to the manifest diagonality and isotropy of the energy-momentum tensor, there are really only two independent components of $\mathbb{T}_{\mu \nu}$, i.e. $\mathbb{T}_{\tau \tau}$ and $\mathbb{T}_{i i}$ for a given $i$. Alternatively one can consider the $\mathbb{T}_{\tau \tau}$ component and the trace $\mathbb{T}^{\mu}_{\mu}$, as by definition
\begin{eqnarray}
\nonumber \mathbb{T}_{\mu}^{\mu} &=& g^{\mu \nu} \mathbb{T}_{\mu \nu} = C^{-2} \mathbb{T}_{\tau \tau} - 3 C^{-2} \mathbb{T}_{i i}
\end{eqnarray}
and then
\begin{eqnarray}\label{TracePressure}
\mathbb{T}_{i i} &=& \frac{\mathbb{T}_{\tau \tau}-C^2 \mathbb{T}^{\mu}_{\mu}}{3} \ ,
\end{eqnarray}
where no sum is intended over the index $i$ and we have used the isotropy of $\mathbb{T}_{\mu \nu}$.

It is worth stressing that, as evident from the definition and the expression of the auxiliary tensor, $\mathbb{T}_{\mu \nu}$ depends only on the time coordinate $\tau$ and parametrically on the arbitrary fixed time $\tau_0$. 

We conclude the section by giving the explicit functional form of $\mathbb{T}_{\tau \tau}$ and $\mathbb{T}^{\mu}_{\mu}$. From Eqs.(\ref{VeV2}) and (\ref{BogoliubovCoefficients5}) we have
\begin{eqnarray}\label{EtaEtaFunction}
\nonumber \mathbb{T}_{\tau \tau}\left[\phi_{p;j},\gamma_{p;j} \right] &= & 2 i C^{-2} \sin^2 \theta \sum_{\lambda}\int d^3 p \lbrace |\Xi_p(\tau_0)|^2
\\
\nonumber & \times & \sum_{j=1,2} \left(\phi^*_{p;j} \partial_{\tau} \phi_{p;j}  + \gamma^*_{p;j} \partial_{\tau} \gamma_{p;j}  - \partial_{\tau}\phi^*_{p;j} \phi_{p;j} - \partial_{\tau}\gamma^*_{p;j} \gamma_{p;j} \right)
\\
\nonumber & + & 2i \Im [\Xi_p^* (\tau_0) \Lambda_p (\tau_0) \left(\phi^*_{p;1} \partial_{\tau}\gamma^*_{p;1} - \gamma_{p;1}^*  \partial_{\tau} \phi^*_{p;1}  \right)
\\
\nonumber &- & \Xi^*_p(\tau_0) \Lambda^*_p (\tau_0) \left(\phi^*_{p;2} \partial_{\tau}\gamma^*_{p;2} - \gamma_{p;2}^*  \partial_{\tau} \phi^*_{p;2}  \right) ] \rbrace
\\
\nonumber &-& i C^{-2} \sum_{\lambda}\sum_{j=1,2}\int d^3 p  \left[\phi^*_{p;j} \partial_{\tau} \phi_{p;j}  + \gamma^*_{p;j} \partial_{\tau} \gamma_{p;j}  - \partial_{\tau}\phi^*_{p;j} \phi_{p;j} - \partial_{\tau}\gamma^*_{p;j} \gamma_{p;j} \right]\,, \\
&& \
\end{eqnarray}
and
\begin{eqnarray}\label{TraceFunction}
\nonumber \mathbb{T}_{\mu}^{\mu} \left[\phi_{p;j}, \gamma_{p;j}\right] &=& 4iC^{-3} \sin^2 \theta \sum_{\lambda} \int d^3 p \lbrace -i|\Xi_p(\tau_0)|^2 \sum_{j=1,2} m_j \left(|\phi_{p;j} |^2 - |\gamma_{p;j} |^2 \right)
\\
\nonumber &+& 2i \Im \left[im_2 \Xi_p^* (\tau_0) \Lambda^*_p (\tau_0) \phi^*_{p;2}  \gamma^*_{p;2} -im_1 \Xi_p^*(\tau_0) \Lambda_p(\tau_0) \phi^*_{p;1} \gamma^*_{p;1} \right] \rbrace
\\
 &-& 2C^{-3} \sum_{\lambda} \sum_{j=1,2} m_j \int d^3 p  \left( |\phi_{p;j} |^2 - |\gamma_{p;j} |^2\right) \ .
\end{eqnarray}
Note that both the $\tau \tau$ component (\ref{EtaEtaFunction}) and the trace (\ref{TraceFunction}) can also be seen as functionals of $\lbrace \phi_p,\partial_{\tau}\phi_p \rbrace$. In fact from Eqs.(\ref{DiracEquation5}) we have 
\begin{eqnarray}\label{GammaEquations}
\nonumber \gamma_p (\eta) &=& \frac{i}{p}\left( \partial_{\eta}\phi_p(\eta) + imC(\eta)\phi_p(\eta)\right) \\ \partial_{\eta}\gamma_p(\eta) &=& \frac{i}{p} \left(\partial_{\eta}^2 \phi_p (\eta) + im\partial_{\eta}C(\eta) \phi_p (\eta) + imC(\eta) \partial_{\eta}\phi_p(\eta) \right) \ ,
\end{eqnarray}
which can be substituted in the expressions above for $\mathbb{T}_{\mu\nu}$. From the diagonality of the expectation value it is also straightforward to prove its covariant conservation $\nabla_{\mu}\mathbb{T}^{\mu \nu} = 0$ (see the appendix \ref{ConservAppendix} for the details).

\section{Applications: exponential expansion}

We have shown that independently of the specific scale factor $C$, the energy-momentum tensor associated to the flavor vacuum $\mathbb{T}_{\mu \nu}^{(MIX)}$ satisfies a number of important properties: (i) it is diagonal, (ii) it is covariantly conserved and (iii) depends only on time $\tau$. Then $\mathbb{T}_{\mu \nu}^{(MIX)}$ for the metric \eqref{LineElement} corresponds to the energy-momentum tensor of a perfect fluid with time-dependent energy density and pressure. 
In this section, in order to better understand the properties of $\mathbb{T}_{\mu \nu}^{(MIX)}$, we assume a specific evolution of the scale factor $C $ and compute the corresponding expectation value of the energy-momentum tensor on the flavor vacuum. In doing so, we are neglecting the back-reaction due to the flavor fields, i.e., the modifications induced on the metric by the energy-momentum tensor of Eq. \eqref{VeV}. The computation based on a fixed background metric is undoubtedly an approximation, but it is useful to get an insight into the kind of contribution that emerges from the flavor vacuum. The self-consistent way to deal with $\mathbb{T}_{\mu \nu}$ is to insert Eq. \eqref{VeV}, together with all the relevant matter terms, on the right-hand side of the Einstein equations and then solve simultaneously for the scale factor $C $ and the Dirac modes. This kind of calculation will be performed elsewhere.

\subsection{Positive energy solutions }
Here, in particular, we study the energy momentum-tensor corresponding to an exponential evolution of the scale factor $C(t) = e^{H_0t}$, with $H_0$ a constant with dimensions of mass. The great advantage of so-picked  scale factor is that the corresponding Dirac equation can be solved analytically without resorting to any approximation, allowing for an in-depth analysis of $\mathbb{T}^{(MIX)}_{\mu \nu}$.  Transforming to conformal time we have
\begin{equation}\label{ExponentialScale}
  \tau = -\frac{1}{H_0} e^{-H_0t} \,,\ \ \ \ \ \ \ \ \ C = -\frac{1}{H_0 \tau}  \  .
\end{equation}
Notice that the conformal time $\tau$ is always negative. Inserting Eq. \eqref{ExponentialScale} in Eq. \eqref{PhiEquation} we obtain
\begin{equation}\label{PhiEquation2}
  \tau^2 \partial_{\tau}^2 \phi_p + \left(p^2 \tau^2 + \frac{im}{H_0} + \frac{m^2}{H_0^2} \right) \phi_p = 0 \ ,
\end{equation}
or, introducing the positive variable $s=-p\tau$
\begin{equation}\label{PhiEquation3}
  s^2 \partial_s^2 \phi_p + \left(s^2 + \frac{im}{H_0} + \frac{m^2}{H_0^2} \right) \phi_p = 0 \ .
\end{equation}
This is a Bessel-like equation, whose general solution can be written as
\begin{equation}\label{PhiEquation4}
  \phi_p (s) = s^{\frac{1}{2}} \left(C_1 J_{\nu} (s) + C_2 J_{-\nu} (s) \right) \,,\ \ \ \ \ \ \ \ \ \ \nu = \frac{1}{2}\left(1-\frac{2im}{H_0} \right)
\end{equation}
with $J_{\nu}(s)$ denoting the Bessel function of order $\nu$ and $C_1, C_2$ arbitrary complex constants. In order to specify the solution we need to impose some kind of boundary conditions, which in turn determine the positive energy solutions. We require that the modes of Eq. \eqref{PhiEquation4} be positive energy with respect to $\partial_{\tau}$ at early times, i.e. for $\tau \rightarrow - \infty$ (where $C \rightarrow 0$). With this choice the mass vacuum corresponds to the absence of massive neutrinos at early times. As $\tau \rightarrow -\infty$ the mass terms can be neglected, and equation \eqref{PhiEquation2} becomes
\begin{equation}
  \partial^2_{\tau} \phi_p + p^2 \phi_p = 0 \ .
\end{equation}
The positive energy solution with respect to $\partial_{\tau}$ is evidently $\phi^{+}_p \propto e^{- i p \tau}$, or, in terms of the $s$ variable, $\phi^{+}_p (s) \propto e^{i s}$. We then impose the requirement
$
  \lim_{s \rightarrow + \infty} \phi_p(s) \propto e^{is} \ .
$
Recalling that $s$ is a positive real variable, we can employ the large argument expansion of the Bessel functions \cite{Abramowitz}
\begin{equation}
  J_{\nu} (s) \simeq \sqrt{\frac{2}{\pi s}} \cos \left(s - \frac{\nu \pi}{2} - \frac{\pi}{4} \right) \ \ \ \ for \ \ \ \ \ s \rightarrow + \infty
\end{equation}
In this way, we show that the combination satisfying the requirement is given by
\begin{equation}\label{PhiEquation5}
  \phi_p (s) = N_p s^{\frac{1}{2}} \left(J_{\nu}(s) - ie^{\frac{\pi m}{H_0}} J_{-\nu}(s)\right) \simeq N_p \sqrt{\frac{2}{\pi}} (-i e^{\frac{\pi m}{2H_0}} ) \cosh \left(\frac{\pi m}{H_0} \right) e^{is} \ ,
\end{equation}
where $N_p$ is a normalization constant and the last equivalence holds in the limit $s \rightarrow + \infty$. Inserting Eq.\eqref{PhiEquation5} in the first of the Eqs.\eqref{GammaEquations}, we deduce
\begin{equation}
  \gamma_p (s) = N_p s^{\frac{1}{2}} \left( - i J_{\nu -1} (s) + e^{\frac{\pi m}{H_0}} J_{1 - \nu} (s) \right)
\end{equation}
where we have made use of the differential relations satisfied by the Bessel functions \cite{Abramowitz}. In order to fix $N_p$ we impose the normalization condition \eqref{Normalization2} in the $s \rightarrow + \infty$ limit. Using the large argument expansion once again, we obtain $|N_p|^2 = \frac{1}{32 \pi^2 \cosh^2 \left( \frac{\pi m}{H_0}\right)} e^{- \frac{\pi m}{H_0}} $. Finally, choosing $N_p$ real, we can write the positive energy solutions as
\begin{eqnarray}\label{PositiveSolutions}
 \nonumber  \phi_p (s) &=& \frac{1}{4 \pi }\frac{e^{- \frac{\pi m}{2H_0}}}{\cosh\left(\frac{\pi m}{H_0} \right)} \sqrt{\frac{s}{2}} \left(J_{\nu} (s) - i e^{\frac{\pi m}{H_0}}J_{-\nu} (s) \right) \\
 \gamma_p (s) &=& \frac{1}{4 \pi }\frac{e^{- \frac{\pi m}{2H_0}}}{\cosh\left(\frac{\pi m}{H_0} \right)} \sqrt{\frac{s}{2}} \left(-iJ_{\nu-1} (s) + e^{\frac{\pi m}{H_0}}J_{1-\nu} (s) \right)
\end{eqnarray}
with $\nu = \frac{1}{2} \left(1- \frac{2im}{H_0} \right)$ and $s=-p\tau$.

\subsection{Bogoliubov Coefficients}

\begin{figure}
  \centering
  \includegraphics{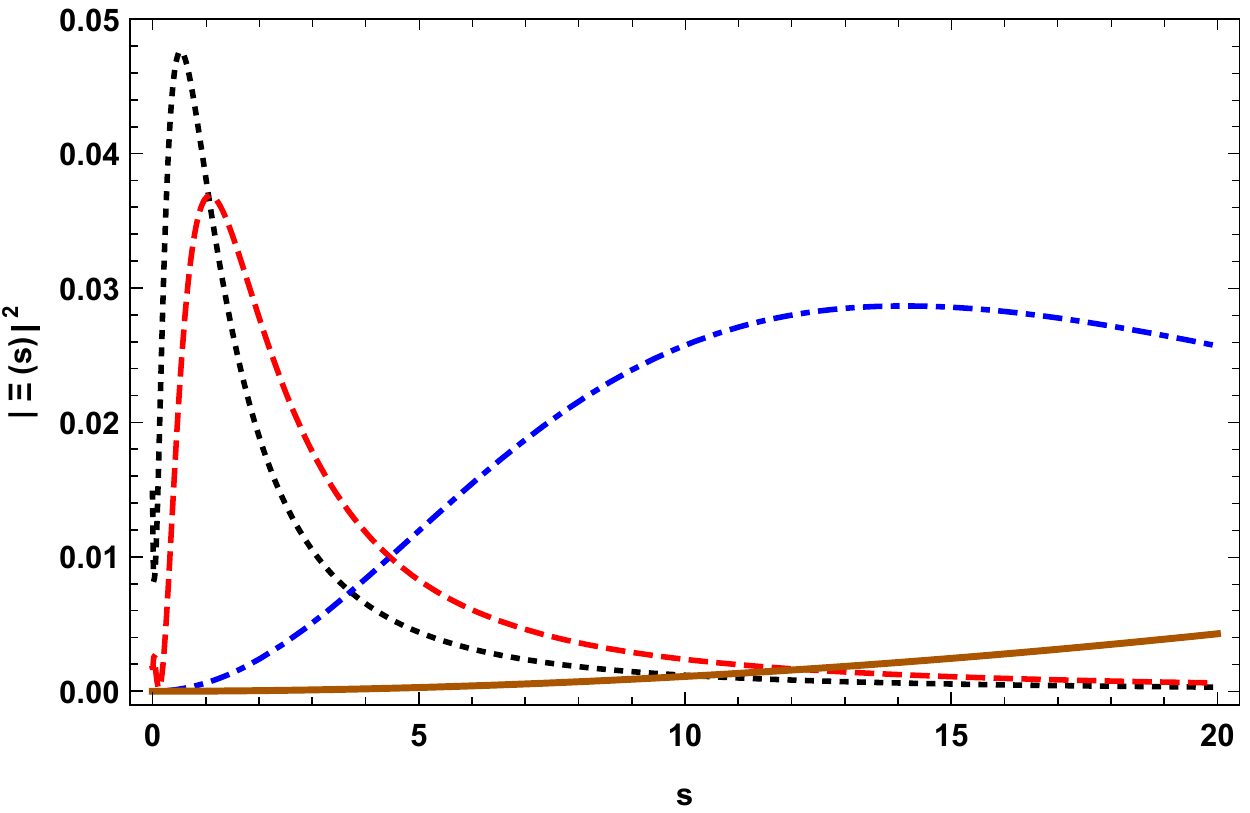}
  \caption{Squared modulus of the Bogoliubov coefficient $|\Xi_p|^2$ as a function of $s$ for sample values of the masses (all in units of $H_0$): (black dotted line) $m_1= 0.7$, $m_2 = 1.4$ , (red dashed line) $m_1 = 1, m_2 = 2$, (blue dotdashed line) $m_1=10 , m_2 =20$, (dark orange solid line), $m_1=100$, $m_2 = 300$. The momentum dependence is implicit in $s = - p \tau$. }
  \label{Figure1}
\end{figure}

We have two sets of solutions $\phi_{p,j}, \gamma_{p,j}$, one for each value of the mass $m_j$, with $ j =1,2$ from the equations \eqref{PositiveSolutions}. The compatibility requirement then implies that each of the $\phi_{p,j}, \gamma_{p,j}$ has the same form for $j=1,2$, except that one has $m_j$ wherever the mass appears, including the function index $\nu_j = \frac{1}{2} \left(1- \frac{2im_j}{H_0} \right)$.
We can compute the Bogoliubov coefficients ($\Lambda_p(\tau), \Xi_p(\tau)$) straight away from Eqs.\eqref{BogoliubovCoefficients5}:
\begin{eqnarray}\label{BogoliubovCoefficients6}
\nonumber \Lambda_p (\tau) &=& \frac{\pi s}{4} \frac{e^{- \frac{\pi}{2H_0}(m_1 + m_2)}}{\cosh \left(\frac{\pi m_1}{H_0} \right)\cosh \left(\frac{\pi m_2}{H_0} \right)} \Bigg \lbrace J_{\nu_2}^*(s) J_{\nu_1} (s) + J^*_{\nu_2 - 1} (s) J_{\nu_1 -1} (s) + i e^{\frac{\pi m_1}{H_0}} \left[ J^*_{\nu_2 - 1} (s) J_{1-\nu_1}(s) - J^*_{\nu_2} (s) J_{-\nu_1}(s)\right] \\
\nonumber &+& i e^{\frac{\pi m_2}{H_0}} \left[ J^*_{-\nu_2}(s) J_{\nu_1}(s) - J^*_{1-\nu_2}(s)J_{\nu_1 -1}(s)\right] + e^{\frac{\pi}{H_0}(m_1 + m_2)} \left[J^*_{-\nu_2}(s) J_{-\nu_1}(s) + J^*_{1-\nu_2}(s) J_{1-\nu_1}(s) \right] \Bigg \rbrace \\
\nonumber \Xi_p (\tau) &=& \frac{\pi s}{4} \frac{e^{- \frac{\pi}{2H_0}(m_1 + m_2)}}{\cosh \left(\frac{\pi m_1}{H_0} \right)\cosh \left(\frac{\pi m_2}{H_0} \right)} \Bigg \lbrace i\left[J^*_{\nu_1}(s)J^*_{\nu_2-1}(s) - J^*_{\nu_1 -1} (s) J^*_{\nu_2} (s) \right] + e^{\frac{\pi m_2}{H_0}} \left[ J^*_{\nu_1} (s) J^*_{1-\nu_2}(s) + J^*_{\nu_1-1} (s) J^*_{-\nu_2}(s)\right] \\
 &-& e^{\frac{\pi m_1}{H_0}} \left[ J^*_{-\nu_1}(s) J^*_{\nu_2-1}(s) + J^*_{1-\nu_1}(s)J^*_{\nu_2}(s)\right] + i e^{\frac{\pi}{H_0}(m_1 + m_2)} \left[J^*_{-\nu_1}(s) J^*_{1-\nu_2}(s) - J^*_{1-\nu_1}(s) J^*_{-\nu_2}(s) \right] \Bigg \rbrace
\end{eqnarray}
To give a flavor of the behaviour of the Bogoliubov coefficients we have plotted $|\Xi_p (s)|^2$ as a function of $s$ for sample values of $m_1, m_2$ in figure (\ref{Figure1}).

We shall be particularly interested in the late time expression of the Bogoliubov coefficients, namely for $s \rightarrow 0^{+}$ (the corresponding limit is $\tau \rightarrow 0^{-}$, or $t \rightarrow + \infty$). For its determination we make use of the small argument expansion of the Bessel functions $J_{\nu}(s) \simeq \left( \frac{s}{2}\right)^{\nu} \frac{1}{\Gamma(1+\nu)}$ with $\Gamma$ denoting the Euler gamma function. From the Eqs.\eqref{BogoliubovCoefficients6}, it is easy to find that at the leading order for $s \rightarrow 0^{+}$ one has
\begin{eqnarray}\label{Bogoliubovcoefficients7}
\nonumber \Lambda_p (s) &\simeq& \frac{\pi}{2} \frac{e^{-\frac{\pi}{2H_0}(m_1 + m_2)}}{\cosh\left(\frac{\pi m_1}{H_0} \right) \cosh\left(\frac{\pi m_2}{H_0} \right)} \left[\frac{e^{i \left(\frac{m_2 - m_1}{H_0}\right) \log \left(\frac{s}{2} \right)}}{\Gamma^*(\nu_2)\Gamma(\nu_1)} + e^{\frac{\pi}{H_0}(m_1 + m_2)}\frac{e^{-i \left(\frac{m_2 - m_1}{H_0}\right) \log \left(\frac{s}{2} \right)}}{\Gamma^*(1-\nu_2)\Gamma(1-\nu_1)}\right]
\\\nonumber
\\
\Xi_p (s) &\simeq& \frac{\pi}{2} \frac{e^{-\frac{\pi}{2H_0}(m_1 + m_2)}}{\cosh\left(\frac{\pi m_1}{H_0} \right) \cosh\left(\frac{\pi m_2}{H_0} \right)} \left[\frac{e^{\frac{\pi m_2}{H_0}} e^{-i \left(\frac{m_2 - m_1}{H_0}\right) \log \left(\frac{s}{2} \right)}}{\Gamma^*(\nu_1)\Gamma^*(1-\nu_2)} - \frac{e^{\frac{\pi m_1}{H_0}}e^{i \left(\frac{m_2 - m_1}{H_0}\right) \log \left(\frac{s}{2} \right)}}{\Gamma^*(1-\nu_1)\Gamma^*(\nu_2)}\right]
\end{eqnarray}

\subsection{Explicit form of the energy-momentum tensor}

For the explicit calculation it is convenient to refer to the splitting of Eq.\eqref{VeV2} and compute $\mathbb{T}_{\mu \nu}^{(MIX)}$ and $\mathbb{T}_{\mu \nu}^{(N)}$ separately. Moreover, keeping in mind the results of the previous sections, it is sufficient to compute the $\tau\tau$ component and the trace in order to fully determine $\mathbb{T}_{\mu \nu}$. Inserting the solutions \eqref{PositiveSolutions} in the Eq.\eqref{EtaEtaFunction} we find, after a lengthy but straightforward calculation:
\begin{eqnarray}\label{EtaEtaFunction2}
\nonumber \mathbb{T}^{(MIX)}_{\tau \tau} (\tau) &=& i \sin^2 \theta \sum_{\lambda} \int d^3 p |\Xi_p (\tau_0)|^2 \left( \frac{H^2_0 p^2 \tau^3}{16 \pi^2} \right) \sum_{j=1,2} \frac{e^{-\frac{\pi m_j}{H_0}}}{\cosh^2 \left(\frac{\pi m_j}{H_0} \right)} \Bigg \lbrace \bigg[ 2(J^*_{\nu_j}J_{\nu_j-1}-J^*_{\nu_j-1}J_{\nu}) + \frac{\nu_j^*-\nu_j}{s} |J_{\nu_j}|^2 \\
\nonumber &+& \frac{\nu_j - \nu_j^*}{s}|J_{\nu_j-1}|^2 \bigg] + i e^{\frac{\pi m_j}{H_0}} \bigg[2 \left(J^*_{\nu_j}J_{1-\nu_j} + J^*_{-\nu_j}J_{\nu_j-1} + J^*_{\nu_j-1}J_{-\nu_j} + J^*_{1-\nu_j} J_{\nu_j} \right) + \frac{\nu_j -\nu_j^*}{s} J^*_{\nu_j}J_{-\nu_j} \\
\nonumber &+& \frac{\nu_j^* - \nu_j}{s}J^*_{-\nu_j}J_{\nu_j} + \frac{\nu_j-\nu_j^*}{s} J^*_{\nu_j-1}J_{1-\nu_j} + \frac{\nu_j^*-\nu_j}{s} J^*_{1-\nu_j}J_{\nu_j-1} \bigg] + e^{\frac{2\pi m_j}{H_0}} \bigg[2(J^*_{1-\nu_j}J_{-\nu_j}-J^*_{-\nu_j}J_{1-\nu_j}) \\
\nonumber &+& \frac{\nu_j^* - \nu_j}{s}|J_{-\nu_j}|^2 + \frac{\nu_j -\nu_j^*}{s}|J_{1-\nu_j}|^2 \bigg] \Bigg \rbrace  \\
\nonumber &+& \frac{i}{2} \sin^2 \theta \sum_{\lambda} \int d^3 p \Bigg \lbrace \Xi^*_p (\tau_0) \Lambda_p (\tau_0) \left(\frac{H_0^2 p^2 \tau^3 e^{-\frac{\pi m_1}{H_0}}}{8 \pi^2 \cosh^2 \left(\frac{\pi m_1}{H_0} \right)} \right) \bigg[ \left(-i (J^{*}_{\nu_1})^2 - i (J^*_{\nu_1-1})^2 + i \frac{2\nu^*_1 - 1}{s} J^*_{\nu_1} J^*_{\nu_1 -1}\right) \\
\nonumber &+& e^{\frac{\pi m_1}{H_0}} \left( 2(J^*_{\nu_1}J^*_{-\nu_1}-J^*_{\nu_1-1}J^*_{1-\nu_1}) + \frac{2\nu_1^* - 1}{s}J^*_{\nu_1}J^*_{1-\nu_1}+ \frac{1-2\nu_1^*}{s}J^*_{-\nu_1}J^*_{\nu_1-1}\right)  \\
\nonumber &+&  i e^{\frac{2 \pi m_1}{H_0}} \left( (J^*_{-\nu_1})^2 + (J^*_{1-\nu_1})^2 + \frac{2\nu_1^* -1}{s} J^*_{-\nu_1}J^*_{1-\nu_1} \right) \bigg] - c.c. \Bigg \rbrace \\
\nonumber &-& \frac{i}{2}\sin^2 \theta \sum_{\lambda} \int d^3 p \Bigg \lbrace \Xi^*_p (\tau_0) \Lambda^*_p (\tau_0) \left(\frac{H_0^2 p^2 \tau^3 e^{-\frac{\pi m_2}{H_0}}}{8 \pi^2 \cosh^2 \left(\frac{\pi m_2}{H_0} \right)} \right) \bigg[ \left(-i (J^{*}_{\nu_2})^2 - i (J^*_{\nu_2-1})^2 + i \frac{2\nu^*_2 - 1}{s} J^*_{\nu_2} J^*_{\nu_2 -1}\right) \\
\nonumber &+& e^{\frac{\pi m_2}{H_0}} \left( 2(J^*_{\nu_2}J^*_{-\nu_2}-J^*_{\nu_2-1}J^*_{1-\nu_2}) + \frac{2\nu_2^* - 1}{s}J^*_{\nu_2}J^*_{1-\nu_2}+ \frac{1-2\nu_2^*}{s}J^*_{-\nu_2}J^*_{\nu_2-1}\right)  \\
\nonumber &+&  i e^{\frac{2 \pi m_2}{H_0}} \left( (J^*_{-\nu_2})^2 + (J^*_{1-\nu_2})^2 + \frac{2\nu_2^* -1}{s} J^*_{-\nu_2}J^*_{1-\nu_2} \right) \bigg] - c.c. \Bigg \rbrace \\ 
\ \ \ \ \ \ \ \
\end{eqnarray}
In the above equation we have left the Bogoliubov coefficients implicit and the argument of the Bessel functions $s=-p\tau$ has been omitted for a better visualization. We recall that the helicity sum is over $\lambda = \pm 1$ and that in general $\tau \neq \tau_0$. An analogous calculation can be performed for the trace:
\begin{eqnarray}\label{TraceFunction2}
\nonumber \mathbb{T}^{\mu (MIX)}_{\mu} &=& i \sin^2 \theta \sum_{\lambda} \int d^3 p |\Xi_p (\tau_0)|^2 \left(\frac{i H_0^3 \tau^3 s}{8 \pi^2}\right) \sum_{j=1,2} \left( \frac{m_j e^{\frac{-\pi m_j}{H_0}}}{\cosh^2 \left( \frac{\pi m_j}{H_0} \right)} \right) \Bigg \lbrace |J_{\nu_j}|^2 - |J_{\nu_j-1}|^2 \\
\nonumber &+& i e^{\frac{\pi m_j}{H_0}} \left(J^*_{-\nu_j}J_{\nu_j} - J^*_{\nu_j}J_{-\nu_j} + J^*_{1-\nu_j}J_{\nu_j-1} - J^*_{\nu_j-1}J_{1-\nu_j} \right) + e^{\frac{2 \pi m_j}{H_0}} \left(|J_{-\nu_j}|^2 - |J_{1-\nu_j}|^2 \right) \Bigg \rbrace \\
\nonumber &+& \frac{i}{2} \sin^2 \theta \sum_{\lambda} \int d^3 p \Bigg \lbrace \Xi_p^*(\tau_0) \Lambda_p (\tau_0) \left( \frac{im_1 s H_0^3 \tau^3 e^{-\frac{\pi m_1}{H_0}}}{4 \pi^2 \cosh^2 \left(\frac{\pi m_1}{H_0} \right)} \right)\bigg[ i J^*_{\nu_1}J^*_{\nu_1 - 1} + e^{\frac{\pi m_1}{H_0}} J^*_{\nu_1}J^*_{1-\nu_1}  \\
\nonumber &-& e^{\frac{\pi m_1}{H_0}}J^*_{-\nu_1}J^*_{\nu_1-1} + ie^{\frac{2 \pi m_1}{H_0}} J^*_{-\nu_1} J^*_{1-\nu_1} \bigg] - c.c \Bigg \rbrace \\
\nonumber &-& \frac{i}{2} \sin^2 \theta \sum_{\lambda} \int d^3 p \Bigg \lbrace \Xi_p^*(\tau_0) \Lambda^*_p (\tau_0) \left( \frac{im_2 s H_0^3 \tau^3 e^{-\frac{\pi m_2}{H_0}}}{4 \pi^2 \cosh^2 \left(\frac{\pi m_2}{H_0} \right)} \right)\bigg[ i J^*_{\nu_2}J^*_{\nu_2 - 1} + e^{\frac{\pi m_2}{H_0}} J^*_{\nu_2}J^*_{1-\nu_2}  \\
 &-& e^{\frac{\pi m_2}{H_0}}J^*_{-\nu_2}J^*_{\nu_2-1} + ie^{\frac{2 \pi m_2}{H_0}} J^*_{-\nu_2} J^*_{1-\nu_2} \bigg] - c.c \Bigg \rbrace
\ \ \ \ \ \ \ .
\end{eqnarray}
 We are particularly interested in the late time ($\tau \rightarrow 0^{-}$) expression of the above equations. According to the definition \eqref{VeV}, these represent the contribution of the flavor vacuum state, defined at an earlier time $\tau_0 < \tau $, to the energy-momentum tensor at late times. We then perform the small argument expansion $J_{\nu}(-p\tau) \simeq \left(\frac{-p\tau}{2}\right)^{\nu} \frac{1}{\Gamma(1+\nu)}$ for all the Bessel functions appearing in Eqs. \eqref{EtaEtaFunction2} and \eqref{TraceFunction2}, and keep only the terms of lowest order in the variable $\tau$. At order $\tau$, the $\tau \tau$ component is found to be
\begin{eqnarray}\label{EtaEtaFunction4}
  \nonumber \mathbb{T}_{\tau \tau}^{(MIX) (1)} (\tau) &\simeq& i \sin^2 \theta \sum_{\lambda} \int d^3 p |\Xi_p(\tau_0)|^2 \left(i \frac{H_0 \tau}{2 \pi^3} \right) \sum_{j=1,2} m_j \tanh \left(\frac{\pi m_j}{H_0} \right) \\ 
 \nonumber &+& \frac{i}{2} \sin^2 \theta \sum_{\lambda} \int d^3 p \left[\Xi^*_p (\tau_0) \Lambda_p(\tau_0) \left(\frac{-im_1 H_0 \tau}{2 \pi^3 \cosh \left( \frac{\pi m_1}{H_0} \right)} \right) - c.c. \right] \\
  &-& \frac{i}{2} \sin^2 \theta \sum_{\lambda} \int d^3 p \left[\Xi^*_p (\tau_0) \Lambda^*_p(\tau_0) \left(\frac{-im_2 H_0 \tau}{2 \pi^3 \cosh \left( \frac{\pi m_2}{H_0} \right)} \right) - c.c. \right]  \  .
\end{eqnarray}
The corresponding lowest order in the trace is $\propto \tau^3$. To see why this is the case, recall that by definition
\begin{equation}
  \mathbb{T}_{\mu}^{\mu} = C^{-2} \mathbb{T}_{\tau \tau} - C^{-2} \sum_{l=1}^{3}\mathbb{T}_{ll} =H_0^2 \tau^2 \mathbb{T}_{\tau \tau} - H_0^2 \tau^2 \sum_{l=1}^{3}\mathbb{T}_{ll} \ ,
\end{equation}
so that in correspondence with $\mathbb{T}_{\tau \tau} \propto \tau$ one has $\mathbb{T}^{\mu}_{\mu} \propto \tau^3$. To this order the trace is
\begin{eqnarray}\label{TraceFunction4}
\nonumber \mathbb{T}^{\mu \ (MIX) \ (1)}_{\mu} &\simeq& i \sin^2 \theta \sum_{\lambda} \int d^3 p |\Xi_p(\tau_0)|^2 \left(\frac{iH_0^3\tau^3}{2\pi^3} \right) \sum_{j=1,2} m_j \tanh \left(\frac{\pi m_j}{H_0} \right) \\ 
 \nonumber &+& \frac{i}{2} \sin^2 \theta \sum_{\lambda} \int d^3 p \left[\Xi^*_p (\tau_0) \Lambda_p(\tau_0) \left(\frac{-im_1 H_0^3 \tau^3}{2 \pi^3 \cosh \left( \frac{\pi m_1}{H_0} \right)} \right) - c.c. \right] \\
  &-& \frac{i}{2} \sin^2 \theta \sum_{\lambda} \int d^3 p \left[\Xi^*_p (\tau_0) \Lambda^*_p(\tau_0) \left(\frac{-im_2 H_0^3 \tau^3}{2 \pi^3 \cosh \left( \frac{\pi m_2}{H_0} \right)} \right) - c.c. \right]  \  .
\end{eqnarray}

Inserting Eqs. \eqref{EtaEtaFunction4} and \eqref{TraceFunction4} in Eq. \eqref{TracePressure}, we can deduce the important result
$
  \mathbb{T}_{ii}^{(MIX)(1)} = 0 \ \,,   \forall i   \ .
$
In other words, at first order in $\tau$, the spatial components of the energy-momentum tensor vanish. Then the equation of state reads, at lowest order in $\tau$:
\begin{equation}\label{EquationOfState}
  w^{(1)} = \frac{\mathbb{T}^{i \ (MIX) (1)}_i }{\mathbb{T}^{\tau \ (MIX) (1)}_{\tau}} = 0 \ ,
\end{equation}
i. e., \emph{the energy-momentum tensor associated to the flavor vacuum satisfies, at late times ($\tau \rightarrow 0^{-}$), the equation of state of a pressure-less perfect fluid}. It is important to stress that this result does not depend on the value of the momentum integrals, and therefore is independent of any regularization. Moreover, as it is evident from \eqref{EquationOfState} it holds for any choice of the reference time $\tau_0$.

\subsection{Regularization}

All the momentum integrals in the expression for the energy-momentum tensor, both for the general case and the for the late time approximation, are to be performed over the whole of $\mathbb{R}^3$ and are formally divergent. To extract a finite result, we need some form of regularization. The most immediate way to regularize the momentum integrations is to introduce an ultraviolet momentum cut-off $\mathcal{K}$. Usually, the cut-off is chosen in correspondence with a 'new physics' energy scale, beyond which the 'low energy' quantum field theory description breaks down. In our case, $\mathcal{K}$ ought to be related to the scale at which the semiclassical approximation breaks down, i. e., at the quantum gravity scale $\mathcal{K} \simeq M_{pl}$ which is of the order of the Planck mass $M_{pl}$.
However, we can expect the cut-off to be at a much lower scale for what concerns the proper mixing term. Indeed, at very high energies, the mass difference between the neutrino states becomes negligible, and the oscillation frequency $\propto \frac{1}{p}$ approaches zero, implying that there is no oscillation. The same result also holds in quantum field theory, where the mixing Bogoliubov coefficient $\Xi_p$ generally approaches zero at high energies (therefore yielding no contribution to the energy-momentum tensor.). The heuristic argument above provides a physical reason to adopt the cut-off regularization and also justifies the adoption of a cut-off scale $\mathcal{K} \ll M_{pl}$, at least for what concerns the mixing.

Before proceeding we need to clarify that the cut-off must be imposed upon the comoving momentum
$
  \pmb{p}_{PHYS} = \frac{\pmb{p}}{C } \ .
$
rather than the mode label $p$.

The imposition of a cut-off $\mathcal{K}_0$ on $p_{PHYS}$ then translates into a sort of 'comoving' cutoff for the mode label $p$
\begin{equation}\label{TrueCutOff}
  p_{CUTOFF} = \mathcal{K}_0 C = \frac{-\mathcal{K}_0}{H_0 \tau} = \mathcal{K}(\tau) \ ,
\end{equation}
which is strictly positive (recall that $\tau < 0$).
For the late time energy-momentum tensor, in the approximation in which $\tau_0 < \tau$ is also at late times $\tau_0 \rightarrow 0^{-}$, we can give a simple analytical form of the regularized integrals. Performing the integrals in Eqs. \eqref{EtaEtaFunction4} and \eqref{TraceFunction4}, with the comoving cut-off of Eq. \eqref{TrueCutOff}, we obtain
\begin{eqnarray}\label{EtaEtaFunction7}
 \nonumber  \mathbb{T}_{\tau \tau}^{(MIX) \ (1) } &=&  \frac{-\sin^2 \theta H_0 \tau \mathcal{K}^3(\tau)}{3 \pi^2} \Bigg[ \frac{e^{\frac{\pi (m_2-m_1)}{H_0}} + e^{-\frac{\pi (m_2-m_1)}{H_0}}}{\cosh \left(\frac{\pi m_1}{H_0} \right) \cosh \left(\frac{\pi m_2}{H_0} \right)} \left(m_1 \tanh \left( \frac{\pi m_1}{H_0} \right) + m_2 \tanh \left( \frac{\pi m_2}{H_0}\right) \right) \\ 
 \nonumber &-& 2 \frac{m_1 \tanh \left(\frac{\pi m_2}{H_0} \right)}{\cosh^2 \left(\frac{\pi m_1}{H_0} \right)} - 2 \frac{m_2 \tanh \left(\frac{\pi m_1}{H_0} \right)}{\cosh^2 \left(\frac{\pi m_2}{H_0} \right)} \Bigg] + \sin^2 \theta H_0 \tau \mathcal{K}^3 (\tau) \Bigg[ \left(\frac{m_1 \tanh \left( \frac{\pi m_1}{H_0}\right)+ m_2 \tanh \left( \frac{\pi m_2}{H_0}\right)}{\cosh^2 \left( \frac{\pi m_1}{H_0}\right) \cosh^2 \left(\frac{\pi m_2}{H_0}\right)} \right) \\ 
 \nonumber &\times& \left(\frac{1}{\Gamma(\nu_1) \Gamma^*(\nu_2)} \right)^2 \left(\frac{1}{3 + 2i \frac{m_2-m_1}{H_0}} \right) \left(\frac{-\mathcal{K}(\tau) \tau_0}{2} \right)^{\frac{2i(m_2-m_1)}{H_0}} + c.c. \Bigg] \\
 \nonumber &+& i \sin^2 \theta \mathcal{K}^3 (\tau) \Bigg \lbrace \Bigg[ \left(\frac{-im_1 H_0 \tau e^{\frac{-\pi m_1}{H_0}}}{2\cosh^3 \left(\frac{\pi m_1}{H_0} \right) \cosh^2 \left(\frac{\pi m_2}{H_0}\right)} \right)\left(\frac{1}{\Gamma(\nu_1) \Gamma^*(\nu_2)} \right)^2 \left(\frac{1}{3 + 2i \frac{m_2-m_1}{H_0}} \right) \left(\frac{-\mathcal{K}(\tau) \tau_0}{2} \right)^{\frac{2i(m_2-m_1)}{H_0}} \\ 
 \nonumber &-& \left(\frac{-im_1 H_0 \tau e^{\frac{-\pi m_1}{H_0}}}{2\cosh^3 \left(\frac{\pi m_1}{H_0} \right) \cosh^2 \left(\frac{\pi m_2}{H_0}\right)} \right)\left(\frac{1}{\Gamma^*(\nu_1) \Gamma(\nu_2)} \right)^2 \left(\frac{1}{3 - 2i \frac{m_2-m_1}{H_0}} \right) \left(\frac{-\mathcal{K}(\tau) \tau_0}{2} \right)^{\frac{-2i(m_2-m_1)}{H_0}} \Bigg] - c.c. \Bigg \rbrace \\ 
 \nonumber &-& i \sin^2 \theta \mathcal{K}^3 (\tau) \Bigg \lbrace \Bigg[ \left(\frac{-im_2 H_0 \tau e^{\frac{-\pi m_2}{H_0}}}{2\cosh^3 \left(\frac{\pi m_2}{H_0} \right) \cosh^2 \left(\frac{\pi m_1}{H_0}\right)} \right)\left(\frac{1}{\Gamma(\nu_1) \Gamma^*(\nu_2)} \right)^2 \left(\frac{1}{3 + 2i \frac{m_2-m_1}{H_0}} \right) \left(\frac{-\mathcal{K}(\tau) \tau_0}{2} \right)^{\frac{2i(m_2-m_1)}{H_0}} \\ 
 \nonumber &-& \left(\frac{-im_2 H_0 \tau e^{\frac{-\pi m_2}{H_0}}}{2\cosh^3 \left(\frac{\pi m_2}{H_0} \right) \cosh^2 \left(\frac{\pi m_1}{H_0}\right)} \right)\left(\frac{1}{\Gamma^*(\nu_1) \Gamma(\nu_2)} \right)^2 \left(\frac{1}{3 - 2i \frac{m_2-m_1}{H_0}} \right) \left(\frac{-\mathcal{K}(\tau) \tau_0}{2} \right)^{\frac{-2i(m_2-m_1)}{H_0}} \Bigg] - c.c. \Bigg \rbrace \\
 \ \ \ \ \ \ .
\end{eqnarray}
The trace can be deduced immediately from Eq. \eqref{EtaEtaFunction7}  by simply multiplying by the factor $C^{-2} $, since the pressure is zero for the late time energy momentum tensor (see Eq. \eqref{EquationOfState}).

\begin{figure}
  \centering
  \includegraphics{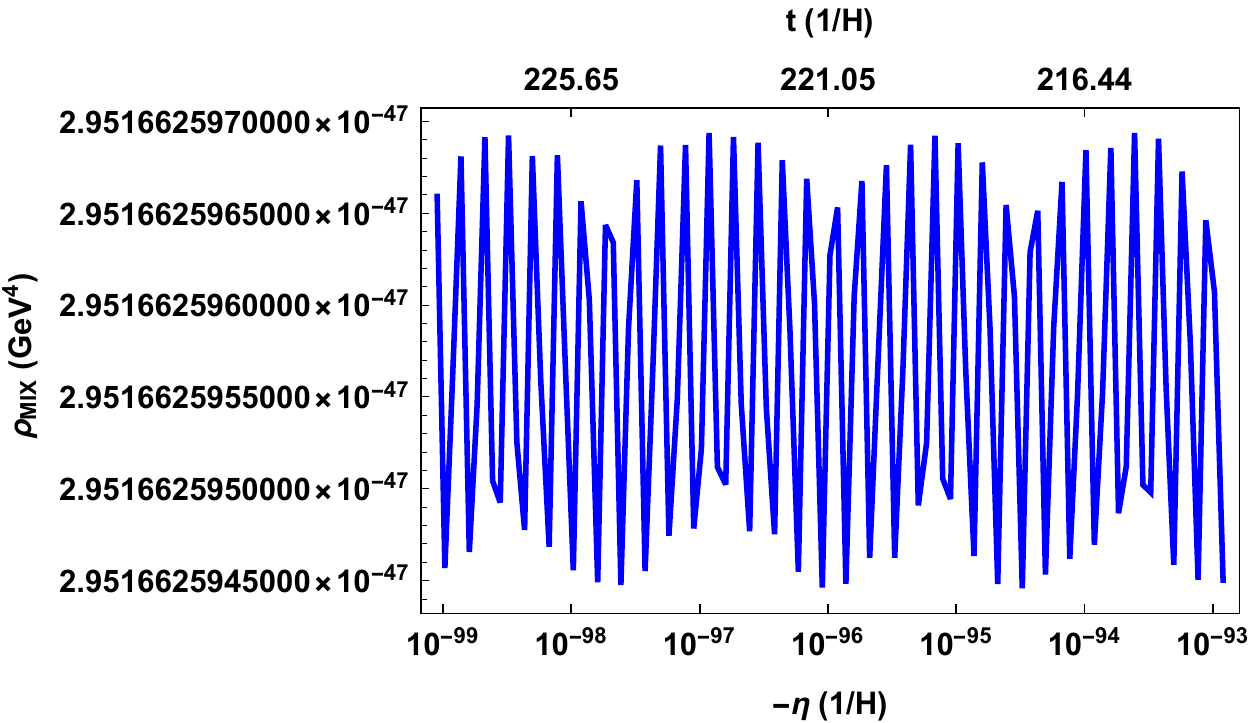}
  \caption{ Logarithmic scale plot of the energy density $\rho_{MIX} = \mathbb{T}^{\tau (MIX)}_{\tau}$ from Eq. \eqref{EtaEtaFunction7} as a function of conformal time $\tau$ for sample values of the parameters. The corresponding coordinate time $t$ is reported above. We have used a cut-off $\mathcal{K}_0 = 246 \ \mathrm{GeV}$ of the order of the electroweak scale, neutrino masses $m_1 = 15.25 H_0 \ , m_2 = 22.25 H_0$ and the expansion rate $H_0 = 10^{-3} \mathrm{eV} $.}
  \label{Figure2}
\end{figure}

A visual indication of the physical content of Eq. \eqref{EtaEtaFunction7} is provided in Figure (2), where the mixing energy density $\mathbb{T}^{\tau \ (MIX)}_{\tau}$ from equation \eqref{EtaEtaFunction7} is plotted against conformal time $\tau$ for sample values of the parameters. We notice that $\rho_{MIX}$, for the parameters and the range considered in Figure (2), is nearly constant, except for tiny oscillations of relative magnitude $\delta \rho_{MIX}/\rho_{MIX} \simeq 10^{-9}$. The oscillations come from the imaginary exponentials in Eq. \eqref{EtaEtaFunction7}. Such a simple evolution pattern can be expected to disappear when other time ranges are considered, and the full expression from Eq. \eqref{EtaEtaFunction2} is employed, entailing a much more intricate time evolution. The results obtained in this work indicate that the vacuum energy associate to neutrino mixing in curved space might represent a dark matter component. 
Notice that for a momentum cutoff of the order of the electroweak scale $\mathcal{K}_0 = 246 \ \mathrm{GeV}$, for neutrino masses $m_1, m_2$ such that $\Delta m^2_{12} \simeq 10^{-5} \mathrm{eV}^2$ and for a value of $H_0 \simeq 10^{-3} \mathrm{eV}$, we obtain an energy density $\mathbb{T}^{\tau (MIX)}_{\tau} $ which is compatible with the upper bound on the dark matter content of the universe. A value of $H_0 \simeq 10^{-3} \mathrm{eV}$ might have been reached during the very early phases of the Universe, i.e. during the first second after the Big Bang.

\section{Conclusions}

In this paper we have considered the quantum field theory of fermion mixing on curved spacetime and we have computed the expectation value $\mathbb{T}_{\mu \nu}^{(MIX)}$ of the energy-momentum tensor of fermion fields on the flavor vacuum in a flat FLRW background. $\mathbb{T}_{\mu \nu}^{(MIX)}$ behaves as an effective energy-momentum tensor, that satisfies the Bianchi identities (i.e. its divergence with respect to one of the two indices vanishes) and therefore can be employed as a regular source for the Einstein equations. In particular, it turns out that $\mathbb{T}_{\mu \nu}^{(MIX)}$ can be interpreted as the energy-momentum tensor of a barotropic fluid.

In this picture, therefore, quantum effects of the fermion fields can be associated, at classical level, to additional barotropic fluids whose thermodynamical properties depend on the geometry of the spacetime other than the specific features of the the fermion field itself. In light of the matter field interpretation of the dark components of the Universe this result seems quite encouraging, as it links directly quantum effects to effective classical fluids. Thus our results show explicitly that  there can be a bridge between quantum properties of matter and  dark matter/energy. However, the presence of a connection between classical fluids and quantum effect is not automatically sufficient alone to conclude that quantum effects are the prime cause of dark matter/energy. In any case, our findings provide an indication that the vacuum energy associated to neutrino mixing can represent a component of dark matter.

Here we are specifically interested in determining the form that the energy--momentum tensor, associated to flavor vacuum, assumes in cosmological FLRW metrics. In order to grasp the behaviour of the energy--momentum tensor of the flavor vacuum, we have then assumed a specific form of the metric. In doing so, we have clearly considered that the term due to the flavor vacuum is not the dominant source of the Einstein equations, which are instead determined by some other source that forces the specific form of the metric.

In particular, considering a de Sitter underlying metric, we have derived the components of $\mathbb{T}_{\mu \nu}^{(MIX)}$  exactly using Bessel functions.  It turns out that at first order in the time parameter only $\mathbb{T}_{\tau \tau}^{(MIX)}$ is different form zero. Hence, in this case the mixing of fermion fields can be associated to a zero pressure (dust or cold dark matter) fluid. We remark that the choice of the de Sitter metric is dictated, primarily, by the fact that exact analytical solutions of the Dirac equation can be found in this context. 
The actual identification of such a contribution as a dark matter component requires that the study be conducted on metrics that are adequate to the description of galaxies. 
Such an analysis will be performed in a forthcoming paper, where we will also consider the contribution due to the flavor vacuum as the dominant source in the Einstein equations. 
Nevertheless, the results presented here clearly hint at a dark--matter--like behaviour of the flavor vacuum in curved background.


   
\appendix


\section{Orthonormality and completeness of the solution of Dirac equation}

Knowing that the determinant is $g =- C^8 $, we can easily compute the scalar product between the solutions (\ref{Solutions}):
\begin{eqnarray}\label{ScalarProduct2}
 \nonumber  (u_{\pmb{p},\lambda},u_{\pmb{q},\lambda'})_{\tau} &=& \int_{\Sigma_{\tau}}d^3x C^3 u^{\dagger}_{\pmb{p}, \lambda} u_{\pmb{q},\lambda'} \\ \nonumber &=& \int_{\Sigma_{\tau}}d^3x C^3 e^{-i(\pmb{p}-\pmb{q})\cdot \pmb{x}} \xi^{\dagger}_{\lambda} \xi_{\lambda'} \left(f^*_{p}f_q + \lambda \lambda'g^*_p g_q\right) \\
 &=&(2\pi)^3 C^{3} \delta^{3}(\pmb{p}-\pmb{q})\delta_{\lambda,\lambda'}\left(|f_p|^2 + |g_p|^2 \right) \ .
\end{eqnarray}
In the last step we have used the definition of the Dirac delta function and the orthonormality of the helicity bispinors. In a similar way it is easy to show that $(u_{\pmb{p},\lambda},v_{\pmb{q},\lambda'})_{\tau}=0$,
$(v_{\pmb{p},\lambda},u_{\pmb{q},\lambda'})_{\tau}=0$ and
\begin{equation}\label{ScalarProduct3}
  (v_{\pmb{p},\lambda},v_{\pmb{q},\lambda'})_{\tau} = (2\pi)^3 C^{3} \delta^{3}(\pmb{p}-\pmb{q})\delta_{\lambda,\lambda'}\left(|f_p|^2 + |g_p|^2 \right) \ .
\end{equation}
Eqs.(\ref{ScalarProduct2}) and (\ref{ScalarProduct3}) suggest that we adopt the following normalization:
\begin{equation}\label{Normalization}
  |f_p|^2+|g_p|^2 =\frac{1}{\left(2\pi C\right)^3}
\end{equation}
in order that the orthonormality of the solutions is satisfied. The set of solutions (\ref{Solutions}) is complete:
\begin{eqnarray}\label{Completeness}
\nonumber \sum_{\lambda} \left(u_{\pmb{p},\lambda}u^{\dagger}_{\pmb{p},\lambda} + v_{\pmb{p},\lambda}v^{\dagger}_{\pmb{p},\lambda}\right) &=&  \sum_{\lambda} \left[ \begin{pmatrix} f_p \xi_{\lambda}
 \\ g_p \lambda \xi_{\lambda}\end{pmatrix}\begin{pmatrix} f^*_p \xi^{\dagger}_{\lambda} & g^*_p \lambda \xi^{\dagger}_{\lambda} \end{pmatrix} + \begin{pmatrix} g^*_p \xi_{\lambda} \\ -f^*_p \lambda \xi_{\lambda}\end{pmatrix}\begin{pmatrix} g_p \xi^{\dagger}_{\lambda} & -f_p \lambda \xi^{\dagger}_{\lambda} \end{pmatrix} \right] \\ \nonumber
&=& \sum_{\lambda} \xi_{\lambda}\xi^{\dagger}_{\lambda} \begin{pmatrix} |f_p|^2 + |g_p|^2 & 0 \\ 0 & |f_p|^2 + |g_p|^2 \end{pmatrix}
\\
&=&  \frac{1}{\left(2\pi C\right)^3} \begin{pmatrix} \mathbb{I} & 0 \\ 0 & \mathbb{I} \end{pmatrix} \ .
\end{eqnarray}
In the last step we have used the completeness of the helicity basis $\sum_{\lambda}\xi_{\lambda}\xi^{\dagger}_{\lambda} = \mathbb{I}$, with $\mathbb{I}$ the $2\times2$ identity matrix, and the normalization condition (\ref{Normalization}).

\section{Properties of helicity eigenbispinors}

We prove that for each $\lambda=\pm 1$ the quantity $\xi^{\dagger}_{\lambda}(\hat{p})\sigma_i \xi_{\lambda}(\hat{p})$ is an odd function of $p_i$, for $i=1,2,3$. In particular, $\xi^{\dagger}_{\lambda}(\hat{p})\sigma_i \xi_{\lambda}(\hat{p})$ changes sign when the momentum is reversed $\pmb{p}\rightarrow - \pmb{p}$.

\textbf{Proof:} The statement can be proven by direct calculation. For the $\lambda=+1$ case, we have
\begin{eqnarray}
\nonumber \xi^{\dagger}_{+}\sigma_1\xi_{+} &=& \begin{pmatrix} e^{i\frac{\phi_p}{2}}\cos(\frac{\theta_p}{2}) & e^{-i\frac{\phi_p}{2}}\sin(\frac{\theta_p}{2}) \end{pmatrix} \begin{pmatrix} 0 & 1\\ 1 & 0 \end{pmatrix}\begin{pmatrix} e^{-i\frac{\phi_p}{2}}\cos(\frac{\theta_p}{2}) \\ e^{i\frac{\phi_p}{2}}\sin(\frac{\theta_p}{2}) \end{pmatrix}
 \\ &=& \cos(\frac{\theta_p}{2})\sin(\frac{\theta_p}{2})\left( e^{i\phi_p}+ e^{-i\phi_p} \right) = \sin(\theta_p)\cos(\phi_p) = \frac{p_1}{p}
 \\\nonumber
 \\ \nonumber \xi^{\dagger}_{+}\sigma_2\xi_{+} &=& \begin{pmatrix} e^{i\frac{\phi_p}{2}}\cos(\frac{\theta_p}{2}) & e^{-i\frac{\phi_p}{2}}\sin(\frac{\theta_p}{2}) \end{pmatrix} \begin{pmatrix} 0 & -i\\ i & 0 \end{pmatrix}\begin{pmatrix} e^{-i\frac{\phi_p}{2}}\cos(\frac{\theta_p}{2}) \\ e^{i\frac{\phi_p}{2}}\sin(\frac{\theta_p}{2}) \end{pmatrix} \\
 &=& \cos(\frac{\theta_p}{2})\sin(\frac{\theta_p}{2})\left( ie^{-i\phi_p}-i e^{i\phi_p} \right) = \sin(\theta_p)\sin(\phi_p) = \frac{p_2}{p}
 \\\nonumber
 \\
 \nonumber \xi^{\dagger}_{+}\sigma_3\xi_{+} &=& \begin{pmatrix} e^{i\frac{\phi_p}{2}}\cos(\frac{\theta_p}{2}) & e^{-i\frac{\phi_p}{2}}\sin(\frac{\theta_p}{2}) \end{pmatrix} \begin{pmatrix} 1 & 0\\ 0 & -1 \end{pmatrix}\begin{pmatrix} e^{-i\frac{\phi_p}{2}}\cos(\frac{\theta_p}{2}) \\ e^{i\frac{\phi_p}{2}}\sin(\frac{\theta_p}{2}) \end{pmatrix} \\
 &=& \cos^2(\frac{\theta_p}{2})-\sin^2(\frac{\theta_p}{2}) = \cos(\theta_p) = \frac{p_3}{p}
\end{eqnarray}
similarly, for the $\lambda = -1$ case we have
\begin{eqnarray}
\nonumber \xi^{\dagger}_{-}\sigma_1\xi_{-} &=& \begin{pmatrix} e^{i\frac{\phi_p}{2}}\sin(\frac{\theta_p}{2}) & -e^{-i\frac{\phi_p}{2}}\cos(\frac{\theta_p}{2}) \end{pmatrix} \begin{pmatrix} 0 & 1\\ 1 & 0 \end{pmatrix}\begin{pmatrix} e^{-i\frac{\phi_p}{2}}\sin(\frac{\theta_p}{2}) \\ -e^{i\frac{\phi_p}{2}}\cos(\frac{\theta_p}{2}) \end{pmatrix}
 \\ &=& - \cos(\frac{\theta_p}{2})\sin(\frac{\theta_p}{2})\left( e^{i\phi_p}+ e^{-i\phi_p} \right) = -\sin(\theta_p)\cos(\phi_p) = \frac{-p_1}{p}
 \\\nonumber
 \\
\nonumber  \xi^{\dagger}_{-}\sigma_2\xi_{-} &=& \begin{pmatrix} e^{i\frac{\phi_p}{2}}\sin(\frac{\theta_p}{2}) & -e^{-i\frac{\phi_p}{2}}\cos(\frac{\theta_p}{2}) \end{pmatrix} \begin{pmatrix} 0 & -i\\ i & 0 \end{pmatrix}\begin{pmatrix} e^{-i\frac{\phi_p}{2}}\sin(\frac{\theta_p}{2}) \\ -e^{i\frac{\phi_p}{2}}\cos(\frac{\theta_p}{2}) \end{pmatrix} \\
 &=& -\cos(\frac{\theta_p}{2})\sin(\frac{\theta_p}{2})\left( ie^{-i\phi_p}-i e^{i\phi_p} \right) = -\sin(\theta_p)\sin(\phi_p) = \frac{-p_2}{p}
 \\\nonumber
 \\
 \nonumber \xi^{\dagger}_{-}\sigma_3\xi_{-} &=& \begin{pmatrix} e^{i\frac{\phi_p}{2}}\sin(\frac{\theta_p}{2}) & -e^{-i\frac{\phi_p}{2}}\cos(\frac{\theta_p}{2}) \end{pmatrix} \begin{pmatrix} 1 & 0\\ 0 & -1 \end{pmatrix}\begin{pmatrix} e^{-i\frac{\phi_p}{2}}\sin(\frac{\theta_p}{2}) \\ -e^{i\frac{\phi_p}{2}}\cos(\frac{\theta_p}{2}) \end{pmatrix} \\
 &=& -\cos^2(\frac{\theta_p}{2})+\sin^2(\frac{\theta_p}{2}) = -\cos(\theta_p) = \frac{-p_3}{p} \ .
\end{eqnarray}
This result can be summarized in the equation
\begin{equation}\label{Claim}
  \xi^{\dagger}_{\lambda} \pmb{\sigma} \xi_{\lambda} = \frac{\lambda\pmb{p}}{p} \ .
\end{equation}

\section{The Auxiliary Tensor}

This Appendix is devoted to exploring the properties of the auxiliary tensor
\begin{equation}{\label{AuxiliaryTensor}}
  L_{\mu \nu}(A,B)=\bar{A} \tilde{\gamma}_{\mu}(x) D_{\nu}B + \bar{A} \tilde{\gamma}_{\nu}(x) D_{\mu}B - D_{\mu}\bar{A} \tilde{\gamma}_{\nu}(x) B - D_{\nu}\bar{A} \tilde{\gamma}_{\mu}(x) B \ .
\end{equation}
 The first property is obvious from the definition: $L_{\mu \nu}(A,B) = L_{\nu \mu}(A,B)$ for any $A,B$. The second property can be seen at once by comparing the definition (\ref{AuxiliaryTensor}) with the form of the energy-momentum tensor (\ref{EnergyPulseTensor1}). Since $T_{\mu \nu}$ is real, we immediately deduce that $L_{\mu \nu}(A,A)$ is pure imaginary for each solution $A$. The third property can be easily deduced by tracing the definition:
\begin{equation}\label{TraceProperty}
  L_{\mu}^{\mu}(A,B) = g^{\mu \nu}L_{\mu \nu}(A,B) = 2\left( \bar{A} \tilde{\gamma}^{\mu}(x) D_{\mu}B - D_{\mu}\bar{A} \tilde{\gamma}_{\mu}(x) B\right) = -4im\bar{A}B
\end{equation}
where we have used the Dirac equation and its adjoint in the last step.
Next we give a more explicit form for $L_{\tau \tau} (A,B)$. Given the expression of the $\tilde{\gamma}_ {\mu}$ matrices we have
\begin{eqnarray}\label{EtaEtaProperty}
  \nonumber L_{\tau\tau} (A,B)&=& 2\left( \bar{A} \tilde{\gamma}_{\tau}(x) D_{\tau}B - D_{\tau}\bar{A} \tilde{\gamma}_{\tau}(x) B\right) = 2C \left( \bar{A} \gamma^0 \partial_{\tau}B-\partial_{\tau}\bar{A}\gamma^0 B \right) \\ &=& 2C \left(A^{\dagger}\partial_{\tau}B - \partial_{\tau}A^{\dagger}B\right) \ .
\end{eqnarray}
In particular we are interested in the auxiliary tensors $L_{\tau \tau}(u(v),u(v))$ and the traces $L^{\mu}_{\mu}(u(v),u(v))$computed on the solutions (\ref{Solutions}). For the traces we have
\begin{eqnarray}
  \nonumber L^{\mu}_{\mu}(u_{\pmb{p},\lambda},u_{\pmb{p},\lambda}) &=& -4im\bar{u}_{\pmb{p},\lambda}u_{\pmb{p},\lambda} = -4imu^{\dagger}_{\pmb{p},\lambda} \gamma^0 u_{\pmb{p},\lambda}
  \\
  \nonumber &=& -4im \begin{pmatrix}f^*_p \xi^{\dagger}_{\lambda} & g^*_p \lambda\xi^{\dagger}_{\lambda} \end{pmatrix} \begin{pmatrix} \mathbb{I} & 0 \\ 0 & -\mathbb{I} \end{pmatrix} \begin{pmatrix} f_p \xi_{\lambda} \\ g_p  \lambda\xi_{\lambda} \end{pmatrix} \\
   &=& -4im(|f_p|^2 -|g_p|^2) = -4imC^{-3} \left( |\phi_p|^2 - |\gamma_p|^2\right)
   \\
   \nonumber
   \\
   \nonumber L^{\mu}_{\mu}(u_{\pmb{p},\lambda},v_{\pmb{p},\lambda}) &=& -4im\bar{u}_{\pmb{p},\lambda}v_{\pmb{p},\lambda} = -4imu^{\dagger}_{\pmb{p},\lambda} \gamma^0 v_{\pmb{p},\lambda} \\
  \nonumber &=& -4im \begin{pmatrix}f^*_p \xi^{\dagger}_{\lambda} & g^*_p \lambda\xi^{\dagger}_{\lambda} \end{pmatrix} \begin{pmatrix} \mathbb{I} & 0 \\ 0 & -\mathbb{I} \end{pmatrix} \begin{pmatrix} g_p^* \xi_{\lambda} \\ -f^*_p  \lambda \xi_{\lambda} \end{pmatrix} \\
   &=& -8imf^*_p g^*_p = -8imC^{-3} \phi^*_p \gamma^*_p
   \\
   \nonumber
   \\
     \nonumber L^{\mu}_{\mu}(v_{\pmb{p},\lambda},u_{\pmb{p},\lambda}) &=& -4im\bar{v}_{\pmb{p},\lambda}u_{\pmb{p},\lambda} = -4imv^{\dagger}_{\pmb{p},\lambda} \gamma^0 u_{\pmb{p},\lambda} \\
  \nonumber &=& -4im \begin{pmatrix}g_p \xi^{\dagger}_{\lambda} & -f_p \lambda\xi^{\dagger}_{\lambda} \end{pmatrix} \begin{pmatrix} \mathbb{I} & 0 \\ 0 & -\mathbb{I} \end{pmatrix} \begin{pmatrix} f_p \xi_{\lambda} \\ g_p  \lambda \xi_{\lambda} \end{pmatrix} \\
   &=& -8imf_p g_p = -8imC^{-3} \phi_p \gamma_p = -\left(L^{\mu}_{\mu}(u_{\pmb{p},\lambda},v_{\pmb{p},\lambda}) \right)^*
   \\
   \nonumber
   \\
   \nonumber L^{\mu}_{\mu}(v_{\pmb{p},\lambda},v_{\pmb{p},\lambda}) &=& -4im\bar{v}_{\pmb{p},\lambda}v_{\pmb{p},\lambda} = -4imv^{\dagger}_{\pmb{p},\lambda} \gamma^0 v_{\pmb{p},\lambda} \\
  \nonumber &=& -4im \begin{pmatrix}g_p \xi^{\dagger}_{\lambda} & -f_p \lambda\xi^{\dagger}_{\lambda} \end{pmatrix} \begin{pmatrix} \mathbb{I} & 0 \\ 0 & -\mathbb{I} \end{pmatrix} \begin{pmatrix} g^*_p \xi_{\lambda} \\ -f^*_p  \lambda \xi_{\lambda} \end{pmatrix} \\
  \nonumber &=& -4im(|g_p|^2 -|f_p|^2) = -4imC^{-3} \left( |\gamma_p|^2 - |\phi_p|^2\right) \\ &=& - L^{\mu}_{\mu}(u_{\pmb{p},\lambda},u_{\pmb{p},\lambda}) \ .
\end{eqnarray}
We then compute the $\tau \tau$ components
\begin{eqnarray}
\nonumber L_{\tau \tau}(u_{\pmb{p},\lambda},u_{\pmb{p},\lambda}) &=& 2C \left[u^{\dagger}_{\pmb{p},\lambda} \partial_{\tau}u_{\pmb{p},\lambda} - \partial_{\tau}u^{\dagger}_{\pmb{p},\lambda} u_{\pmb{p},\lambda} \right] \\
\nonumber &=& 2C \left[\begin{pmatrix} f_p^* \xi^{\dagger}_{\lambda} & g_p^*\lambda \xi^{\dagger}_{\lambda} \end{pmatrix} \begin{pmatrix} \partial_{\tau}f_p \xi_{\lambda} \\ \partial_{\tau}g_p \lambda \xi_{\lambda} \end{pmatrix} - \begin{pmatrix} \partial_{\tau}f_p^* \xi^{\dagger}_{\lambda} & \partial_{\tau}g_p^*\lambda \xi^{\dagger}_{\lambda} \end{pmatrix} \begin{pmatrix} f_p \xi_{\lambda} \\ g_p \lambda \xi_{\lambda} \end{pmatrix} \right]
\\
\nonumber &=& 2C \left[ f^*_p \partial_{\tau}f_p - \partial_{\tau}f^*_p f_p + g^*_p \partial_{\tau} g_p - \partial_{\tau}g^*_p g_p\right] \\
&=& 2C^{-2} \left[\phi^*_p \partial_{\tau}\phi_p - \partial_{\tau}\phi^*_p \phi_p + \gamma^*_p \partial_{\tau} \gamma_p - \partial_{\tau}\gamma^*_p \gamma_p \right] \\
\nonumber \\
\nonumber L_{\tau \tau}(u_{\pmb{p},\lambda},v_{\pmb{p},\lambda}) &=& 2C \left[u^{\dagger}_{\pmb{p},\lambda} \partial_{\tau}v_{\pmb{p},\lambda} - \partial_{\tau}u^{\dagger}_{\pmb{p},\lambda} v_{\pmb{p},\lambda} \right] \\
\nonumber &=& 2C \left[\begin{pmatrix} f_p^* \xi^{\dagger}_{\lambda} & g_p^*\lambda \xi^{\dagger}_{\lambda} \end{pmatrix} \begin{pmatrix} \partial_{\tau}g_p^* \xi_{\lambda} \\ -\partial_{\tau}f_p^* \lambda \xi_{\lambda} \end{pmatrix} - \begin{pmatrix} \partial_{\tau}f_p^* \xi^{\dagger}_{\lambda} & \partial_{\tau}g_p^*\lambda \xi^{\dagger}_{\lambda} \end{pmatrix} \begin{pmatrix} g_p^* \xi_{\lambda} \\ -f^*_p \lambda \xi_{\lambda} \end{pmatrix} \right] \\
\nonumber &=& 2C \left[ f^*_p \partial_{\tau}g_p^* - \partial_{\tau}f^*_p g_p^* - g^*_p \partial_{\tau} f_p^* + \partial_{\tau}g^*_p f^*_p\right] \\
&=& 4C^{-2} \left[\phi^*_p \partial_{\tau}\gamma^*_p - \gamma^*_p \partial_{\tau} \phi_p^*\right] \\
\nonumber \\
L_{\tau \tau}(v_{\pmb{p},\lambda},u_{\pmb{p},\lambda})&=& - \left(L_{\tau \tau}(u_{\pmb{p},\lambda},v_{\pmb{p},\lambda}) \right)^* \\
\nonumber \\
L_{\tau \tau}(v_{\pmb{p},\lambda},v_{\pmb{p},\lambda}) & =& - L_{\tau \tau}(u_{\pmb{p},\lambda},u_{\pmb{p},\lambda}) \ .
\end{eqnarray}
Notice that as a consequence $L_{\tau \tau}(u_{\pmb{p},\lambda},v_{\pmb{p},\lambda}) = 0$ if and only if $\gamma_p \propto \phi_p$. This is the case, for instance, in Minkowski spacetime, where $\phi_p \propto \gamma_p \propto e^{-i\omega_p t}$. For a general $C $, the proportionality does not hold, and this has dramatic consequences on the vacuum of the theory (eventually leading to particle creation).

In addition one can show that $L_{\tau i} (u(v),u(v))$ is an odd function of the momentum $\pmb{p}$ and that $L_{ij}(u(v),u(v))$ is an odd function with respect to both $p_i$ and $p_j$. The proof requires a lengthy but straightforward calculation:
\begin{eqnarray}\label{OffDiagonal1}
\nonumber L_{\tau i} (u_{\pmb{p},\lambda},u_{\pmb{p},\lambda}) &= & \bar{u}_{\pmb{p},\lambda} \tilde{\gamma}_{\tau} D_{i} u_{\pmb{p},\lambda} + \bar{u}_{\pmb{p},\lambda} \tilde{\gamma}_{i} D_{\tau} u_{\pmb{p},\lambda} -
D_i\bar{u}_{\pmb{p},\lambda} \tilde{\gamma}_{\tau} u_{\pmb{p},\lambda} -
D_{\tau}\bar{u}_{\pmb{p},\lambda} \tilde{\gamma}_{i} u_{\pmb{p},\lambda}
\\
\nonumber
\\
\nonumber&=& C u^{\dagger}_{\pmb{p},\lambda} \left(\partial_{i} + \frac{1}{8} \omega_{i}^{A,B}[\gamma_A,\gamma_B] \right) u_{\pmb{p},\lambda} -
C u^{\dagger}_{\pmb{p},\lambda}\gamma^0 \gamma^i \partial_{\tau} u_{\pmb{p},\lambda} \\
\nonumber &-& C \left(\partial_{i}u^{\dagger}_{\pmb{p},\lambda}\gamma^0 -u^{\dagger}_{\pmb{p},\lambda}\frac{\gamma^0}{8} \omega_{i}^{A,B}[\gamma_A,\gamma_B] \right)\gamma^0 u_{\pmb{p},\lambda} + C \partial_{\tau}u^{\dagger}_{\pmb{p},\lambda}\gamma^0 \gamma^i u_{\pmb{p},\lambda}
\\
\nonumber\\
\nonumber &=& C u^{\dagger}_{\pmb{p},\lambda} \left( i p_i + \frac{\partial_{\tau}C}{2C} \begin{pmatrix} 0 & \sigma_i
\\ \sigma_i & 0 \end{pmatrix}\right)u_{\pmb{p},\lambda} - C u^{\dagger}_{\pmb{p},\lambda}\begin{pmatrix} 0 & \sigma_i \\ \sigma_i & 0 \end{pmatrix}\partial_{\tau}u_{\pmb{p},\lambda} \\
\nonumber & -& C u^{\dagger}_{\pmb{p},\lambda} \left( -i p_i + \frac{\partial_{\tau}C}{2C} \begin{pmatrix} 0 & \sigma_i \\ \sigma_i & 0 \end{pmatrix}\right)u_{\pmb{p},\lambda} + C \partial_{\tau}u^{\dagger}_{\pmb{p},\lambda}\begin{pmatrix} 0 & \sigma_i \\ \sigma_i & 0 \end{pmatrix}u_{\pmb{p},\lambda}
\\
\nonumber
\\
\nonumber &=& 2ip_i C u^{\dagger}_{\pmb{p},\lambda}u_{\pmb{p},\lambda}+
 C \left[\partial_{\tau}u^{\dagger}_{\pmb{p},\lambda}\begin{pmatrix} 0 & \sigma_i \\ \sigma_i & 0 \end{pmatrix}u_{\pmb{p},\lambda} - u^{\dagger}_{\pmb{p},\lambda}\begin{pmatrix} 0 & \sigma_i \\ \sigma_i & 0 \end{pmatrix}\partial_{\tau}u_{\pmb{p},\lambda} \right]
 \\
\nonumber
\\
\nonumber &=& 2i p_i C \left[|f_p|^2 + |g_p|^2\right] + C \lambda\left[\partial_{\tau}f_p^* g_p + \partial_{\tau}g_p^*f_p-f_p^* \partial_{\tau}g_p-g_p^*\partial_{\tau}f_p \right]\xi^{\dagger}_{\lambda}\sigma_i\xi_{\lambda}
\\
\nonumber
\\
 &=& p_i \left\lbrace \frac{i}{\pi^3 C^2 } + \frac{\lambda}{p} \left[ C \left(\partial_{\tau}f_p^* g_p + \partial_{\tau}g_p^*f_p-c.c.\right) \right] \right\rbrace \ .
\end{eqnarray}
In the last step, we have made use of the normalization condition (\ref{Normalization}) and the property

 (\ref{Claim}). As it is evident from Eq.(\ref{OffDiagonal1}), each $\tilde{\gamma}_i$ factor and each spatial derivative $\partial_i$ brings along a factor $p_i$. Then for each $a,b=u_{\pmb{p},\lambda}, v_{\pmb{p},\lambda}$ one has
\begin{equation}\label{OffDiagonal2}
  L_{\tau i} (a,b) = p_i h_{a,b} (p) \ ,
\end{equation}
with $h_{a,b} (p)$ a function of the modulus $p$ alone. In particular
\begin{equation}\label{ZeroMomentum}
L_{\tau i}(u_{\pmb{p},\lambda},v_{\pmb{p},\lambda}) = 0  \ .
\end{equation}
Similarly, for each $a,b=u_{\pmb{p},\lambda}, v_{\pmb{p},\lambda}$
\begin{equation}\label{OffDiagonal3}
  L_{ij}(a,b) = p_i p_j l_{a,b}(p) \ ,
\end{equation}
with $l_{a,b} (p)$ a function of the modulus $p$ alone.

\section{Canonical anticommutation relations}

Demonstration of the canonical anticommutation relations (\ref{AnticommutationRelations})

\begin{eqnarray}
 \nonumber \lbrace \psi_{A}(\tau,\pmb{x}) , \pi_{\psi B} (\tau,\pmb{x'}) \rbrace &=&
  iC^3 \sum_{\lambda,\lambda'}\int d^3p \int d^3 q \left[ \lbrace A_{\pmb{p},\lambda},A^{\dagger}_{\pmb{q},\lambda'} \rbrace\left( u_{\pmb{p},\lambda} u^{\dagger}_{\pmb{q},\lambda'} \right)_{AB} + \lbrace B^{\dagger}_{-\pmb{p},\lambda},B_{-\pmb{q},\lambda'} \rbrace\left( v_{\pmb{p},\lambda} v^{\dagger}_{\pmb{q},\lambda'} \right)_{AB}\right]
  \\
\nonumber &=& iC^{3} \sum_{\lambda} \int d^3 p \left[ \left( u_{\pmb{p},\lambda} (\tau,\pmb {x}) u^{\dagger}_{\pmb{p},\lambda} (\tau,\pmb{x'})\right)_{AB} + \left( v_{\pmb{p},\lambda} (\tau,\pmb {x}) v^{\dagger}_{\pmb{p},\lambda} (\tau,\pmb{x'})\right)_{AB}\right]
\\
\nonumber &=& iC^{3} \sum_{\lambda} \int d^3 p e^{i \pmb{p} \cdot (\pmb{x}-\pmb{x'})} \left[ \left( u_{\pmb{p},\lambda} (\tau,\pmb {0}) u^{\dagger}_{\pmb{p},\lambda} (\tau,\pmb{0})\right)_{AB} + \left( v_{\pmb{p},\lambda} (\tau,\pmb {0}) v^{\dagger}_{\pmb{p},\lambda} (\tau,\pmb{0})\right)_{AB}\right] \\\nonumber
 &=& i C^3 \frac{1}{(2\pi)^3 C^3 } \delta_{AB} \int d^3 p e^{i \pmb{p} \cdot (\pmb{x}-\pmb{x'})}
 \\
 &=&
 i \delta_{AB} \delta^{3} (\pmb{x}-\pmb{x'}) \ .
\end{eqnarray}
In the fourth step we have made use of the completeness relation (Eq.(\ref{Completeness})), writing the $4\times4$ identity matrix explicitly as $\delta_{AB}$.

\section{Conservation of the energy--momentum tensor}
\label{ConservAppendix}

In this appendix we prove explicitly the covariant conservation of the energy momentum tensor associated to the flavor vacuum. We show that
\begin{equation}\label{Divergence}
 \nabla_{\mu} \mathbb{T}^{\mu \nu} = 0
\end{equation}
with $\nabla_{\mu}$ denoting the covariant derivative. There is no need here to distinguish between $\mathbb{T}_{\mu \nu}^{(MIX)}$ and $\mathbb{T}_{\mu \nu}^{(N)}$, since both satisfy eq. \eqref{Divergence}, and so does the full energy-momentum tensor. Preliminarly we derive the connection coefficients for the metric of eq. \eqref{LineElement}. It is easy to see from the definition that the only non vanishing coefficients are 
\begin{equation}\label{Christoffel}
 \Gamma^{\tau}_{\tau \tau} = \Gamma^{\tau}_{ii} = \Gamma^{i}_{\tau i} = \Gamma^{i}_{i \tau} = \frac{\dot{C}}{C} \ .
\end{equation}
Here the dot denotes the derivative with respect to conformal time $\tau$ and no sum is intended over repeated indices. Notice that the coefficients depend only on $\tau$. In terms of the connection coefficients, the covariant divergence reads
\begin{equation}\label{ExplicitDivergence}
 \nabla_{\mu} \mathbb{T}^{\mu \nu} = \partial_{\mu} \mathbb{T}^{\mu \nu} + \Gamma^{\mu}_{\mu \sigma} \mathbb{T}^{\sigma \nu} + \Gamma^{\nu}_{\mu \sigma} \mathbb{T}^{\mu \sigma} \ .
\end{equation}
\begin{itemize}
 \item ($\nu = i$) For $\nu = i$, with $i=1,2,3$ equation \eqref{ExplicitDivergence} becomes
 \begin{equation}
  \nabla_{\mu}\mathbb{T}^{\mu i} = \partial_{\mu} \mathbb{T}^{\mu i} + \Gamma^{\mu}_{\mu \sigma} \mathbb{T}^{\sigma i} + \Gamma^{i}_{\mu \sigma} \mathbb{T}^{\mu \sigma} \ .
 \end{equation}
From the diagonality of $\mathbb{T}^{\mu \nu}$ proved above, we can write
 \begin{equation}\label{Divergence1}
  \nabla_{\mu}\mathbb{T}^{\mu i} =  \partial_{i} \mathbb{T}^{i i} + \sum_{\mu} \Gamma^{\mu}_{\mu i} \mathbb{T}^{i i} + \sum_{\mu} \Gamma^{i}_{\mu \mu} \mathbb{T}^{\mu \mu} \ ,
 \end{equation}
 where no sum is intended over repeated indices and the summations are written out explicitly to avoid confusion. The first term on the right hand side of eq. \eqref{Divergence1} is zero, since $\mathbb{T}^{\mu \nu}$ depends only on $\tau$. Similarly, from eq. \eqref{Christoffel} we know that $\Gamma^{\mu}_{\mu i} = 0 = \Gamma^{i}_{\mu \mu}$ for each $\mu = 0,1,2,3$ and each $i=1,2,3$, so that also the second and the third term on the right hand side of eq. \eqref{Divergence1} vanish. Then
 \begin{equation}
  \nabla_{\mu} \mathbb{T}^{\mu i} = 0 \ \ \ \ \forall i \ .
 \end{equation}
\item ($\nu = \tau$) Only a slightly longer calculation is needed to prove the statement for $\nu = \tau$. Starting from equation \eqref{ExplicitDivergence} we have
\begin{eqnarray}\label{Divergence2}
 \nonumber \nabla_{\mu} \mathbb{T}^{\mu \tau} &=& \partial_{\mu}\mathbb{T}^{\mu \tau} + \Gamma^{\mu}_{\mu \sigma} \Gamma^{\sigma \tau} + \Gamma^{\tau}_{\mu \sigma} \mathbb{T}^{\mu \sigma} \\
 \nonumber &=& \partial_{\tau} \mathbb{T}^{\tau \tau} + \left(\Gamma^{\tau}_{\tau \tau} + \sum_i \Gamma^{i}_{i \tau}\right) \mathbb{T}^{\tau \tau} + \Gamma^{\tau}_{\tau \tau} \mathbb{T}^{\tau \tau} + \sum_{i} \Gamma^{\tau}_{ii} \mathbb{T}^{ii} \\ &=& \partial_{\tau} \mathbb{T}^{\tau \tau} + 5 \Gamma^{\tau}_{\tau \tau} \mathbb{T}^{\tau \tau} + 3 \Gamma^{\tau}_{\tau \tau}\mathbb{T}^{ii} \ ,
\end{eqnarray}
where we have used the diagonality of $\mathbb{T}^{\mu \nu}$ and eqs. \eqref{Christoffel}. For our purposes it is convenient to rewrite eq. \eqref{Divergence2} in terms of $\mathbb{T}_{\tau \tau}$ and the trace $\mathbb{T}^{\mu}_{\mu}$. To this end we employ eqs. \eqref{TracePressure}, \eqref{Christoffel} and lower the indices through the metric of eq. \eqref{LineElement}, obtaining: 
\begin{equation}\label{Divergence3}
 \nabla_{\mu} \mathbb{T}^{\mu \tau} = \partial_{\tau} \left( C^{-4} \mathbb{T}_{\tau \tau} \right) + 6 C^{-5} \dot{C} \mathbb{T}_{\tau \tau} - C^{-3} \dot{C} \mathbb{T}^{\mu}_{\mu} \ .
\end{equation}
From equation \eqref{VeV2}, we know that each of the terms above is the integral of the auxiliary tensor components $L_{\tau \tau}, L^{\mu}_{\mu}$ weighted by $\tau$-independent coefficients (recall that the Bogoliubov coefficients are evaluated at a fixed time $\tau_0$). It is therefore sufficient to prove that
\begin{equation}
  \partial_{\tau} \left( C^{-4} L_{\tau \tau}(a,b) \right) + 6 C^{-5} \dot{C} L_{\tau \tau} (a,b) - C^{-3} \dot{C} L^{\mu}_{\mu} (a,b) = 0
\end{equation}
for each $a,b = u_{\pmb{p},\lambda;j}, v_{\pmb{p},\lambda;j}$, to show that the divergence \eqref{Divergence3} vanishes. To this end we need the second order equations
\begin{eqnarray}\label{Divergence4}
 \nonumber \partial^2_{\tau} \phi_{p;j} &=& - \left(im_j \dot{C} + p^2 + m_j^2 C^2\right) \phi_{p;j} \\
 \partial^2_{\tau} \gamma_{p;j} &=& - \left(-im_j \dot{C} + p^2 + m_j^2 C^2\right) \gamma_{p;j} \ .
\end{eqnarray}
The first of this equations is simply eq. \eqref{PhiEquation}, and the second can be likewise deduced from the system \eqref{DiracEquation5}. For $L_{\mu \nu} (u_{\pmb{p},\lambda;j},u_{\pmb{p},\lambda;j})$ we have, using the properties of the auxiliary tensor 
\begin{eqnarray*}
 && \partial_{\tau} \left(C^{-4} L_{\tau \tau} (u_{\pmb{p},\lambda;j},u_{\pmb{p},\lambda;j})  \right) + 6 C^{-5} \dot{C} L_{\tau \tau} (u_{\pmb{p},\lambda;j},u_{\pmb{p},\lambda;j}) - C^{-3}\dot{C} L_{\mu}^{\mu} (u_{\pmb{p},\lambda;j},u_{\pmb{p},\lambda;j}) = \\ &&
 2\partial_{\tau} \left[ C^{-6} \left(\phi_{p;j}^* \overleftrightarrow{\partial_{\tau} }\phi_{p;j} + \gamma_{p;j}^* \overleftrightarrow{\partial_{\tau} }\gamma_{p;j}\right) \right] + 12 C^{-7} \dot{C} \left(\phi_{p;j}^* \overleftrightarrow{\partial_{\tau} }\phi_{p;j} + \gamma_{p;j}^* \overleftrightarrow{\partial_{\tau} }\gamma_{p;j}\right) \\ 
 && + 4 i m_j C^{-6}\dot{C} \left(|\phi_{p;j}|^2 - |\gamma_{p;j}|^2 \right) = \\ && 2C^{-6} \partial_{\tau}\left(\phi_{p;j}^* \overleftrightarrow{\partial_{\tau} }\phi_{p;j} + \gamma_{p;j}^* \overleftrightarrow{\partial_{\tau} }\gamma_{p;j}\right) + 4 i m_j C^{-6}\dot{C} \left(|\phi_{p;j}|^2 - |\gamma_{p;j}|^2 \right) = \\
 && 2C^{-6} \left( \phi_{p;j}^* \partial^2_{\tau} \phi_{p;j} - \phi_{p;j}\partial_{\tau}^2 \phi_{p;j}^* + \gamma_{p;j}^* \partial^2_{\tau} \gamma_{p;j} - \gamma_{p;j}\partial_{\tau}^2 \gamma_{p;j}^* \right) + 4 i m_j C^{-6}\dot{C} \left(|\phi_{p;j}|^2 - |\gamma_{p;j}|^2 \right) = \\
 && 2 C^{-6} \Big \lbrace \phi_{p;j}^* \left[- \left(im_j \dot{C} + p^2 + m^2 C^2 \right) \right] \phi_{p;j} - \phi_{p;j}^* \left[- \left(-im_j \dot{C} + p^2 + m^2 C^2 \right)  \right] \phi_{p;j} \\
 && + \gamma_{p;j}^* \left[- \left(-im_j \dot{C} + p^2 + m^2 C^2 \right) \right] \gamma_{p;j} - \phi_{p;j}^* \left[- \left(im_j \dot{C} + p^2 + m^2 C^2 \right)  \right] \gamma_{p;j} \Big \rbrace \\
 && + 4 i m_j C^{-6}\dot{C} \left(|\phi_{p;j}|^2 - |\gamma_{p;j}|^2 \right) = \\
 && 2C^{-6} \left \lbrace |\phi_{p;j}|^2 \left(-2im_j \dot{C} \right) + |\gamma_{p;j}|^2 \left(2im_j \dot{C} \right) \right \rbrace 4 i m_j C^{-6}\dot{C} \left(|\phi_{p;j}|^2 - |\gamma_{p;j}|^2 \right) = \\
 && -4 i m_j C^{-6}\dot{C} \left(|\phi_{p;j}|^2 - |\gamma_{p;j}|^2 \right) + 4 i m_j C^{-6}\dot{C} \left(|\phi_{p;j}|^2 - |\gamma_{p;j}|^2 \right) = 0 \ .
\end{eqnarray*}
In the fourth step we have used the equations \eqref{Divergence4} and their complex conjugates, and the $\tau$ argument of the functions has been suppressed for notational simplicity. Note that from the properties $L_{\tau \tau} (v_{\pmb{p},\lambda;j},v_{\pmb{p},\lambda;j}) = - L_{\tau \tau} (u_{\pmb{p},\lambda;j},u_{\pmb{p},\lambda;j}) $ and $L_{\mu}^{\mu} (v_{\pmb{p},\lambda;j},v_{\pmb{p},\lambda;j}) = - L_{\mu}^{\mu} (u_{\pmb{p},\lambda;j},u_{\pmb{p},\lambda;j}) $ the same relation is satisfied by the components of $L_{\mu \nu} (v_{\pmb{p},\lambda;j},v_{\pmb{p},\lambda;j})$. Similarly one has 
\begin{eqnarray*}
 && \partial_{\tau} \left( C^{-4} L_{\tau \tau} (u_{\pmb{p},\lambda;j}, v_{\pmb{p},\lambda;j})\right) + 6 C^{-5} \dot{C} L_{\tau \tau} (u_{\pmb{p},\lambda;j}, v_{\pmb{p},\lambda;j}) - C^{-3}\dot{C}L_{\mu}^{\mu} (u_{\pmb{p},\lambda;j}, v_{\pmb{p},\lambda;j}) = \\
 && \partial_{\tau} \left(4 C^{-6} \phi_{p;j}^* \overleftrightarrow{\partial_{\tau}} \gamma_{p;j}^* \right) + 24 C^{-7}\dot{C} \left(4 C^{-6} \phi_{p;j}^* \overleftrightarrow{\partial_{\tau}} \gamma_{p;j}^* \right) + 8 i m_j C^{-6} \dot{C} \phi_{p;j}^* \gamma_{p;j}^* = \\
 && 4C^{-6} \partial_{\tau}\left(C^{-6} \phi_{p;j}^* \overleftrightarrow{\partial_{\tau}} \gamma_{p;j}^* \right) + 8 i m_j C^{-6} \dot{C} \phi_{p;j}^* \gamma_{p;j}^* = \\
 && 4 \left[\phi_{p;j}^* \partial^2_{\tau} \gamma^*_{p;j} - \gamma^{*}_{p;j} \partial^2_{\tau}\phi_{p;j}^* \right] + 8 i m_j C^{-6} \dot{C} \phi_{p;j}^* \gamma_{p;j}^* = \\
 && 4C^{-6} \left \lbrace \phi_{p;j}^* \left[- \left(im_j \dot{C} + p^2 + m^2 C^2 \right) \right] \gamma_{p;j}^* - \phi_{p;j}^* \left[- \left(-im_j \dot{C} + p^2 + m^2 C^2 \right) \right] \gamma_{p;j}^*\right \rbrace \\ 
 && + 8 i m_j C^{-6} \dot{C} \phi_{p;j}^* \gamma_{p;j}^* = \\
 && - 8 i m_j C^{-6} \dot{C} \phi_{p;j}^* \gamma_{p;j}^* + + 8 i m_j C^{-6} \dot{C} \phi_{p;j}^* \gamma_{p;j}^* = 0 \ .
\end{eqnarray*}
In the fourth step we have used the complex conjugates of eqs. \eqref{Divergence4}. Finally, due to the properties $L_{\tau \tau} (v_{\pmb{p},\lambda;j},u_{\pmb{p},\lambda;j}) = - L^*_{\tau \tau} (u_{\pmb{p},\lambda;j},v_{\pmb{p},\lambda;j}) $ and $L_{\mu}^{\mu} (v_{\pmb{p},\lambda;j},u_{\pmb{p},\lambda;j}) = - L_{\mu}^{\mu *} (u_{\pmb{p},\lambda;j},v_{\pmb{p},\lambda;j}) $, the same relation is satisfied by the components of $L_{\mu \nu} (v_{\pmb{p},\lambda;j},u_{\pmb{p},\lambda;j})$. This is sufficient to prove the statement for $\nu = \tau$. 

\end{itemize}

\end{document}